\documentclass[english,10pt,journal,twocolumn,twoside]{IEEEtran}
\usepackage{subfigure}
\usepackage{cite}
\usepackage[T1]{fontenc}
\usepackage[latin9]{inputenc}
\usepackage{stmaryrd}
\usepackage{amssymb}
\usepackage{amsmath}
\usepackage{mathrsfs}
\usepackage{amsfonts}  
\usepackage{subfig}
\usepackage[dvips,final]{graphicx}
\usepackage[dvips]{color}
\usepackage{xcolor}
\usepackage{epsfig,amssymb,amsmath,amsthm}
\usepackage{graphicx,amssymb}
\usepackage{graphicx}
 \usepackage{subfigure}
\usepackage{wrapfig,lipsum,booktabs}
\usepackage{algorithm}
\usepackage{algpseudocode}
\usepackage{cuted}
\usepackage{balance}
\usepackage{tabularx}
\usepackage{booktabs}
\usepackage{stfloats}
\usepackage{amssymb}
\usepackage{dsfont}
\usepackage{esint}
\usepackage{amsmath}
\usepackage{subcaption}

\usepackage{enumitem}
\usepackage{tikz}
\usepackage{caption} 
\usepackage{bm}
\usepackage{comment}
\usepackage{subfig}
\usepackage{stfloats}
\usepackage{pifont}
\usepackage{relsize}\usepackage{subdepth}\allowdisplaybreaks

\newcommand{\beq}{\begin{equation}}
\newcommand{\eeq}{\end{equation}}
\newcommand{\beqn}{\begin{eqnarray}}
\newcommand{\eeqn}{\end{eqnarray}}
\newcommand{\T}{\mathrm T}
\newtheorem{theorem}{Theorem}
\usepackage{xcolor}
\usepackage{babel}
  \theoremstyle{definition}
  
  \theoremstyle{plain}
  
  \newtheorem{lemma}{Lemma}
    \newtheorem{corollary}{Corollary}
  \theoremstyle{plain}
  
\usepackage{babel}
\providecommand{\definitionname}{Definition}

\providecommand{\theoremname}{Theorem}

\ifCLASSINFOpdf
\else
\fi
\begin{document}

%

\title{Model-Based Learning for DOA Estimation with One-Bit Single-Snapshot Sparse Arrays}


\author{\IEEEauthorblockN{Yunqiao~Hu, \emph{Student Member, IEEE}, Shunqiao~Sun, \emph{Senior Member, IEEE}, and Yimin D.\ Zhang, \emph{Fellow, IEEE} }
\thanks{The work of Y. Hu ans S. Sun was
supported in part by U.S. National Science Foundation (NSF) under Grants CCF-2153386, ECCS-2340029, and Alabama Transportation Institute (ATI).  The work of Y.\ D.\ Zhang was supported in part by the NSF under Grant ECCS-2236023 and in part by the Air Force Office
of Scientific Research (AFOSR) under Grant FA9550-23-1-0255. The conference precursor of this work was presented at the 2024 IEEE Annual Asilomar Conference on Signals, Systems, and Computers \cite{Yunqiao_Asilomar_2024}.}
\thanks{Y. Hu and S. Sun are with the Department of Electrical and Computer Engineering, The University of Alabama, Tuscaloosa, AL, USA (emails: yhu62@crimson.ua.edu, shunqiao.sun@ua.edu). }
\thanks{Y.\ D.\ Zhang is with the Department of Electrical and Computer Engineering, Temple University, Philadelphia, PA, USA (email: ydzhang@temple.edu). }\vspace{-0.5em}
}

\maketitle

\markboth{IEEE Journal of Selected Topics in Signal Processing}%
{Hu and et. al.: Model-Based Learning for DOA Estimation with One-Bit Single Snapshot Sparse Arrays}
%





\begin{abstract}
We address the challenging problem of estimating the directions-of-arrival (DOAs) of multiple off-grid signals using a single snapshot of one-bit quantized measurements. Conventional DOA estimation methods face difficulties in tackling this problem effectively. This paper introduces a domain-knowledge-guided learning framework to achieve high-resolution DOA estimation in such a scenario, thus drastically reducing hardware complexity without compromising performance. We first reformulate DOA estimation as a maximum a posteriori (MAP) problem, unifying on-grid and off-grid scenarios under a Laplacian-type sparsity prior to effectively enforce sparsity for both uniform and sparse linear arrays. For off-grid signals, a first-order approximation grid model is embedded into the one-bit signal model. We then reinterpret one-bit sensing as a binary classification task, employing a multivariate Bernoulli likelihood with a logistic link function to enhance stability and estimation accuracy. To resolve the non-convexity inherent in the MAP formulation, we develop augmented algorithmic frameworks based on majorization-minimization principles. Further, we design  model-based inference neural networks by deep unrolling these frameworks, significantly reducing computational complexity while preserving the estimation precision. Extensive simulations demonstrate the robustness of the proposed framework across a wide range of input signal-to-noise ratio values and off-grid deviations. By integrating the unified model-based priors with data-driven learning, this work bridges the gap between theoretical guarantees and practical feasibility in one-bit single-snapshot DOA estimation, offering a scalable, hardware-efficient solution for next-generation radar and communication systems.

\end{abstract}

\vspace{-0.5em}
\section{Introduction}
\IEEEPARstart{D}{irection}-of-arrival (DOA) estimation is fundamentally important in sensor array signal processing with wide application in radar, sonar, navigation, and wireless communications\cite{9585542,9127853,9429942,4350230,7870764}. Commonly used super-resolution DOA estimation algorithms, such as MUltiple SIgnal Classification (MUSIC) \cite{1143830} and estimation of signal parameters via rotational invariant techniques (ESPRIT) \cite{32276}, often assume a  uniform linear array (ULA) configuration, where sensor elements are arranged in a straight line with equal spacing, typically half the signal wavelength. However, in practical applications, achieving higher resolution with large-size ULA requires a substantial number of array elements, significantly increasing hardware costs\cite{7870764}. Furthermore, ULAs are susceptible to mutual coupling effects, which can degrade DOA estimation performance\cite{DOA_mutual_coupling_TAES_2012}. To address this problem, sparse linear arrays (SLAs) have been used over the past few decades to achieve desired apertures with fewer active elements. Some SLA configurations, such as the minimum redundancy array
(MRA)\cite{MRA_1968}, the nested array\cite{5456168}, the co-prime array\cite{5609222}, the generalized coprime array configurations \cite{7012090}, and the maximum inter-element spacing constraint (MISC) \cite{8636158}, 
have been well studied and analyzed in the past decades. Although subspace-based methods, such as MUSIC and weighted subspace fitting \cite{97999}, and covariance matrix-based compressive sensing methods \cite{8472789} can be applied to SLAs exploiting difference coarrays, they require a high number of snapshots to achieve accurate covariance matrix estimation, making them impractical in snapshot-limited scenarios commonly encountered in automotive applications \cite{9429942}. 

In antenna array systems, high-precision, high-sampling-rate analog-to-digital converters (ADCs) are costly and power-intensive. One-bit quantization provides a cost-effective solution, simplifying the sampling hardware and lowering the sampling rates by reducing the number of bits per sample to just one \cite{1039405}. Consequently, one-bit signal processing has drawn significant research interest in many fields. In communications, a novel framework for computing the minimum mean square error (MMSE) channel estimator in one-bit quantized multi-input multi-output (MIMO) systems was developed in \cite{10848316}. In \cite{9444144}, a two-stage detection method for massive MIMO with one-bit ADCs was proposed, which employs Bussgang-based linear receivers and a model-driven neural network,
achieving robust performance at significantly reduced complexity. In \cite{8821391}, a novel one-bit MIMO precoding scheme was proposed using spatial Sigma-Delta modulation to effectively control quantization errors utilizing favorable conditions of a massive MIMO system. In the field of radar signal processing, target parameter estimation based on one-bit quantized measurements has emerged as a prominent research topic \cite{Mixed_ADC_SAM_2024,Arian_One_Bit_Radarconf24,Arian_Mixed_ADC_ICASSP_2025}. A two-stage, tensor-decomposition-based channel-estimation framework for MIMO systems with movable antennas was proposed in \cite{10659325}. Successive transmit-receive sweeps are arranged so that the resulting pilot measurements are stacked into a structured third-order tensor, on which canonical polyadic (CP) decomposition is applied to uniquely recover angle-of-arrival/departure (AoA/AoD) and complex path gains, thereby enabling accurate channel reconstruction for arbitrary antenna positions while reducing pilot overhead far below that of compressed sensing (CS) baselines. The channel state information (CSI) bottleneck in XL-MIMO-OFDM systems with dynamic metasurface antennas is alleviated by microstrip-sequencing the pilots into a fourth-order tensor that admits CP decomposition \cite{10938032}. By exploiting Vandermonde factor matrices, a two-stage closed-form procedure is then used to uniquely estimate AoA/AoD, delays, and gains with training overhead proportional only to the number of multipaths, thereby reducing pilots and improving accuracy relative to MMSE and CS baselines.
In \cite{8822763}, one-bit ADC radar parameter estimation was formulated as a cyclic multivariate weighted least squares problem.
A dimension-reduced generalized approximate message passing approach was introduced in  \cite{9025063}  for one-bit frequency-modulated continuous-wave (FMCW) radar that mitigates high-order harmonic interference.  Multiple sparse recovery algorithms were extended in 
\cite{shang2021weighted} to a one-bit quantization setting, resulting in one-bit sparse learning via iterative minimization (1bSLIM), likelihood-based estimation of sparse parameters (1bLIKES), one-bit sparse iterative covariance-based estimation (1bSPICE), and one-bit iterative adaptive approaches (1bIAA), to generate automotive radar range-Doppler imaging exploiting one-bit quantization.

\vspace{-1.0em}
\subsection{Relevant Works}
Subspace-based methods exploiting one-bit quantized data, along with performance analyses, were presented for both ULAs and SLAs \cite{9585542,7478040,8700277}. It is shown in \cite{9585542} that the identifiability of one-bit DOA estimation with SLAs matches unquantized measurements and a conservative Cram\'{e}r-Rao bound (CRB) for quantized systems is derived. In \cite{8700277}, MUSIC-based DOA estimation leverages an approximate one-bit covariance matrix, derived as a scaled version of the unquantized covariance matrix.
In the past decade, methods leveraging sparse reconstruction \cite{1468495} and CS \cite{1614066} have emerged to address one-bit DOA estimation and sparse sensing, leading to various estimators, such as binary iterative hard thresholding (BIHT) \cite{6418031},  robust one-bit Bayesian compressed sensing (R1BCS) \cite{6963346}. The generalized sparse Bayesian learning (Gr-SBL) algorithm for both single- and multi-snapshot scenarios \cite{8357527} can operate effectively with SLAs. However, all these methods only consider on-grid signals, as they determine signal DOAs by peak searching over a fixed discrete angular spectrum.

For off-grid signals, grid refinement is commonly employed  to maintain estimation accuracy, but it requires denser grids,  increasing computational complexity. In CS-based methods, denser grids also lead to higher correlations among dictionary atoms, thereby degrading algorithm performance. To overcome these challenges, gridless and off-grid estimation methods have been developed to tackle the one-bit off-grid DOA estimation problem. Gridless methods, such as atomic norm minimization (ANM) \cite{zhou2017gridless}, estimate DOAs without grid division but require solving a computationally intensive semidefinite programming (SDP) problem. On the other hand, off-grid methods, such as the off-grid iterative reweighted (OGIR) algorithm \cite{10024794}, alternatively refine on-grid spectrum estimates and grid-gap estimates to enhance estimation accuracy. In \cite{EURASIP_SBL_Off_Grid_DOA_2023}, an improved root SBL (IRSBL) method was developed for off-grid DOA estimation that refines a coarse initial dictionary through polynomial rooting and extends its applicability to nonuniform linear arrays. Off-grid methods generally eliminate the need for dense grid division, making them more flexible and applicable for real-world DOA estimation tasks compared to gridless methods. However, these methods often require hundreds of iterations to converge, with each iteration involving computationally expensive matrix inversion.

A dynamic maximum likelihood estimator was proposed in 
\cite{9573279} and applied to both hybrid analog-digital and fully digital structures, with precise DOA estimates obtainable using either a golden section search method or a gradient search approach. A tensor-based approach was introduced in \cite{10403776} for efficiently estimating both channel state information and target parameters in massive MIMO integrated sensing and communication systems. By parameterizing the communication channel and utilizing a shared training pattern, it formulates the problem as a CP decomposition. The unified algorithm estimates key parameters and incorporates a segment-based training pattern to mitigate beam squint effects, resulting in improved accuracy, resolution, and reduced training overhead.

In recent years, deep learning has emerged as a powerful tool in DOA estimation, due to its ability to learn complex patterns and representations from data \cite{Papageorgiou_TSP_2021}. 
However, classical purely data-driven deep neural network-based methods often lack interpretability, and their generalization ability is largely dependent on the size and diversity of the training dataset. Recently, model-based deep learning has gained significant traction in the signal processing community \cite{10056957,Algorithm_Unrolling_SPM_2021}, including its applications in DOA estimation research \cite{10266765,10289861,10348517,Ruxin_SNN_ICASSP_2025}. For array signal processing, model-based deep learning approaches leverage domain knowledge to combine the strengths of traditional signal processing models with the flexibility of deep learning, thus effectively overcoming the limitations of purely data-driven methods and improving performance. In addition, hybrid approaches combine the interpretability of classical array signal models with the representation power of deep neural networks, thus better addressing limitations of traditional signal processing methods in handling high-dimensional, noisy, or complex data. For example, in \cite{yeganegi2024deep}, the authors successfully demonstrated that a learned iterative shrinkage-thresholding algorithm (LISTA) based framework provides superior DOA estimation capability using a limited number of one-bit quantized measurements. 

\vspace{-0.5em}
\subsection{Contributions}

In this paper, to tackle DOA estimation in a single-snapshot one-bit quantized antenna array, we propose a learning-augmented framework that unifies model-based optimization with data-driven neural networks. At its core, our methodology leverages a maximum a posteriori (MAP) estimation approach, reformulating one-bit DOA estimation as a sparsity-constrained inverse problem. Unlike existing sparse recovery techniques, we adopt a Laplacian-type prior \cite{5256324} to enforce structural sparsity, allowing robust recovery of both on-grid and off-grid signals. For off-grid scenarios, a first-order approximation grid model is integrated into the one-bit signal model, thus bypassing the need for iterative grid refinement and ensuring computational tractability for both ULAs and SLAs. To mitigate the non-convexity inherent in one-bit MAP estimation, we employ a majorization-minimization (MM) approach to construct a surrogate convex function, which serves as a substitute for the original objective function, facilitating more efficient optimization. In addition to using the Gaussian cumulative distribution function (CDF) as the likelihood function in MAP estimation, the problem is also reformulated as a binary classification task, where the likelihood follows a multivariate Bernoulli distribution with a logistic link function. This approach offers advantages such as differentiability, computational efficiency, and seamless integration into deep neural network architectures. Finally, the iterative algorithms derived from this framework are distilled into a deep-unrolled neural network, which reduces computational complexity by unrolling optimization steps into a feedforward architecture while preserving estimation accuracy. This codesign of model-based priors and learning-based inference bridges theoretical guarantees with practical efficiency, enabling real-time deployment in resource-constrained systems.

Our main contributions are summarized as follows.
\begin{itemize}
    \item We formulate one-bit DOA estimation problems using a sparse Bayesian framework and the MAP estimation approach with a Laplacian-type prior to enforce sparsity. 
    A first-order approximation grid model is integrated into the one-bit signal model for off-grid DOA estimation, offering a more practical and easily implementable framework for algorithm derivation.
    \item We introduce an augmented sparse Bayesian framework that reimagines one-bit DOA estimation as a binary classification task. By modeling likelihood as a multivariate Bernoulli distribution with a logistic link function, this approach achieves computational efficiency and differentiability, enabling seamless integration into deep neural networks.  We employ an MM approach to mitigate the non-convexity in the augmented one-bit MAP estimator.
    \item We propose a one-bit DOA inference neural network that synergizes model-based sparse priors with learning-based optimization, harnessing the deep-unrolling paradigm to slash computational complexity while maintaining estimation accuracy. This model-based architecture ensures interpretability inherent to model-driven methods and sustains robustness across broad SNR regimes, bridging the gap between theoretical guarantees and real-world deployment in low-resolution array systems. 
\end{itemize}

In \cite{Yunqiao_Asilomar_2024}, a baseline model-based network was introduced for one-bit off-grid DOA estimation. In this paper, our work further advances the field through multiple key innovations. First, we propose an augmented Bayesian framework that generalizes the likelihood model to a multivariate Bernoulli distribution with a logistic link function, enabling gradient-based optimization and seamless integration with deep neural networks. Second, we employ an MM approach to mitigate the non-convexity inherent in the augmented one-bit MAP estimator. Third, the model-based learning framework is extensively evaluated concerning state-of-the-art approaches.

The rest of the paper is organized as follows. Section \ref{sysmodel and problem fromulation} introduces the one-bit DOA estimation signal model and provides a mathematical formulation for both on-grid and off-grid cases. In Section \ref{SBRI framework}, we derive two algorithm frameworks, Sparse Bayesian Reweighted Iterative (SBRI) and SBRI-X, for one-bit DOA estimation, addressing both on-grid and off-grid DOA estimation problems. Section \ref{learning-based approach} presents model-based neural networks inspired by algorithm unrolling.
Section \ref{Numerical Results} provides numerical results for performance evaluation. Finally, Section \ref{conclusion} concludes the paper.

We adopt the following notation throughout the paper. Scalars are denoted by lowercase lightface letters, vectors by lowercase boldface letters, and matrices by uppercase boldface letters. 
$(\cdot)^{ \rm T}$, $(\cdot)^{ \rm H}$, $(\cdot)^*$ respectively represent the transpose, Hermitian, and conjugation operations.
$\Re(\cdot)$ and $\Im(\cdot)$ respectively return the real and imaginary parts of complex values. 
$\mathrm{sign}\left (\cdot\right)$ denotes the sign function with $\mathrm{sign}\left(x\right)=1$ for $x \ge 0$ and $\mathrm{sign}\left(x\right)=-1$ for $x < 0$, whereas $\mathrm{csgn}\left(\cdot\right)=\mathrm{sign} (\Re\left (\cdot\right)) + j\mathrm{sign}\left(\Im\left(\cdot\right)\right)$ is the complex sign function. Symbol $\odot$ is an element-wise Hadamard product, and  
$\mathrm{diag}(\cdot)$ represents the operation that diagonalizes the input.

\vspace{-0.5em}
\section{System Model}
\label{sysmodel and problem fromulation}

Consider a scenario involving $K$ narrowband, far-field signals, represented as $s_k(t)$ for $k = 1, \dots, K$, impinging on an $M$-element SLA from directions $\bm{\theta} = [\theta_1, \cdots, \theta_K]^{\mathrm T}$. As we focus on single-snapshot DOA estimation, the array signal model with one-bit quantization data is presented as:
\begin{equation}
\begin{aligned}
\overline{\mathbf{y}} &= \mathrm{csgn}\left(\mathbf{y} \right)
= \mathrm{csgn} \big(\mathbf{A}(\bm{\theta})\mathbf{s} + \mathbf{n} \big),
\end{aligned}
\label{signle_snapshot_one_bit_data_model}
\end{equation}
where $\mathbf{y}\in \mathbb{C}^{M\times 1}$ is the received signal vector and $\overline{\mathbf{y}}\in \mathbb{C}^{M\times 1}$ is its one-bit quantized version, $\mathbf{A}(\bm{\theta})=\left[\mathbf{a}(\theta_{1} ),\mathbf{a}(\theta_{2} ), \dots, \mathbf{a}(\theta_{K} )\right]\in \mathbb{C}^{M\times K}$ is the SLA manifold matrix, $\mathbf{s} = [s_1, \cdots, s_K]^{\T} \in \mathbb{C}^{K\times 1}$ is the impinging signal vector, and $\mathbf{n}\in \mathbb{C}^{M\times 1}$ is the complex additive noise vector. 
Each column vector in the array manifold matrix $\mathbf{A}(\bm{\theta})$ is a steering vector, given for the $k$-th signal DOA as:
\begin{align}
\mathbf{a}(\theta_k) = \left[1, e^{j\frac{2\pi d_2}{\lambda}\sin{\theta_k}}, \dots , e^{j\frac{2\pi d_M}{\lambda}\sin{\theta_k}}\right]^{\mathrm T},
\end{align}
where $d_m$, $m=2, \cdots, M$, denote the spacing between the $m$-th SLA element and the first element, and $\lambda$ is wavelength.

\subsection{On-Grid Single-Measurement Vector Model}
To estimate $\bm{\theta}$ from $\left(\ref{signle_snapshot_one_bit_data_model}\right)$, we first construct a classic single-measurement vector model using $N$ grid points: 
\begin{equation}
\overline{\mathbf{y}} = \mathrm{csgn}(\mathbf{\mathcal{A}}\mathbf{x} + \mathbf{n}),
\label{svm_model}
\end{equation}
where $\mathbf{\mathcal{A}} =[\mathbf{a}(\tilde{\theta}_{1}),\mathbf{a}(\tilde{\theta}_{2}), \dots, \mathbf{a}(\tilde{\theta}_{N})]\in \mathbb{C}^{M \times N}$ is the dictionary matrix with $N\gg M> K$, $\mathbf{a}(\tilde{\theta}_{i})\in \mathbb{C}^{M \times 1}$ is a dictionary atom, where the elements of $\bm{\Theta} = \{\tilde{\theta}_{1},\tilde{\theta}_{2}, \dots, \tilde{\theta}_{N}\}\in \mathbb{R}^{N \times 1}$ are obtained from uniform angle-space division, and $\mathbf{x}=\left[x_{1}, x_{2},\dots, x_{N}\right]^{\T}\in \mathbb{R}^{N \times 1}$ defines the dictionary atom coefficients to be estimated. Under this model, the $N$ grid points form the bases for sparse signal representation. When the signal can be expressed by using exactly $K$ bases, 
i.e., $\theta_k \in \bm{\Theta}$, $\forall \ k=1, \cdots, K$, 
this model is referred to as an on-grid model.

\vspace{-1.5em}
\subsection{Off-Grid Single-Measurement Vector Model} 
In practice, however, it is unlikely that the pre-demarcated grid corresponds exactly to all signal DOAs, i.e., $\theta_k \notin \Theta$ for some or all $K$ signals. Assume the grid is sufficiently dense such that the $K$ signals fall within distinct grid regions, and we denote the fixed grid point closest to the true DOA $\theta_k$ as $\tilde{\theta}_{n_{k}}$, $n_{k} \in \left\{1,2,\dots,N\right\}$. Then, the true DOA can be expressed as:
\begin{equation}
\theta_k = \tilde{\theta}_{n_{k}} + (\theta_k - \tilde{\theta}_{n_{k}}),
\end{equation}
where $(\theta_k - \tilde{\theta}_{n_{k}})$ represents the grid-gap between the fixed grid point and the true DOA. Using a first-order Taylor expansion, the steering vector $\mathbf{a}(\theta_k)$ can be approximated as:
\begin{equation}
\hat{\mathbf{a}}(\theta_k) = \mathbf{a}(\tilde{\theta}_{n_{k}}) + \mathbf{b}(\tilde{\theta}_{n_{k}})(\theta_k - \tilde{\theta}_{n_{k}}),
\end{equation}
where $\mathbf{b}(\tilde{\theta}_{n_{k}})=\frac{\partial \bm{a}(\theta)}{\partial \theta}|_{\tilde{\theta}_{n_{k}}}\in \mathbb{C}^{M \times 1}$ is the first-order derivative of $\bm{a}(\theta)$ at $\tilde{\theta}_{n_{k}}$. 
In this case, the dictionary matrix is approximated as $\mathbf{C}(\bm{\beta})=\mathbf{\mathcal{A}} + \mathbf{\mathcal{B}} \mathrm{diag}\left(\bm{\beta}\right)$, where $\mathcal{B} = [\mathbf{b}(\tilde{\theta}_{1}),\mathbf{b}(\tilde{\theta}_{2}), \dots, \mathbf{b}(\tilde{\theta}_{N} )]\in \mathbb{C}^{M \times N}$, and the elements of $\bm{\beta} = \left[\beta_1,\beta_2, \dots, \beta_N\right]^{\T}$ are off-grid gaps, defined as:
\begin{equation}
    \beta_n = \begin{cases}
    \theta_k - \tilde{\theta}_{n_k},\ \textrm{if} \ n=n_k, k\in \left\{ 1,2,...,K \right\},
 \\
  0,\ \hspace{3em} \textrm{otherwise}.
\end{cases}
\end{equation}
By including the approximation error in the measurement noise, the measurement model in (\ref{svm_model}) can be reformulated as:
\begin{equation}
\overline{\mathbf{y}} = \mathrm{csgn} \big(\mathbf{C}(\bm{\beta})\mathbf{x} + \mathbf{n} \big). \label{one_bit_off_grid_single_snpashot_DOA_model}
\end{equation}

\vspace{-1.5em}
\section{Sparse Bayesian Learning Framework for One-Bit DOA Estimation}\label{SBRI framework}

In this section, we introduce the SBRI framework and its augmented variant (SBRI-X) for the estimation of on-grid and off-grid signal DOAs based on one-bit quantized array data.
The core SBRI loop reformulates one-bit MAP inference with a sparsity prior into a majorization-minimization (MM)-based iteratively reweighted least squares (IRLS) procedure. At each iteration, we first construct a surrogate for the quantized CDF likelihood, compute a pseudo-measurement that effectively compensates for the sign nonlinearity, and then solve a closed-form IRLS update for the spectral coefficients. The regularization weight is automatically adjusted based on the norm of the current estimate.

SBRI-X extends this framework by introducing noise offsets and replacing the CDF with a smooth sigmoid function. The sigmoid term is majorized using a second-order (quadratic) upper bound, enabling an alternating optimization scheme that performs two simple updates: an efficient IRLS step for the spectral coefficients and a closed-form update for the noise offsets.

\vspace{-0.5em}
\subsection{One-Bit On-Grid DOA Estimation via SBRI  Framework}\label{original_framework}

We first introduce a probabilistic model to characterize the relationship between unknown atom coefficient vector $\mathbf{x}$ and the observed input $\overline{\mathbf{y}}$. Based on (\ref{one_bit_off_grid_single_snpashot_DOA_model}), the posterior probability of $\mathbf{x}$ under the MAP criterion is primarily determined by the likelihood function $p\left(\overline{\mathbf{y}}|\mathbf{x};\bm{\beta}\right)$ and the prior probability density function (PDF) $p\left(\mathbf{x}\right)$. The denominator of the posterior density is ignored as it is a positive constant that has no impact on the optimization process. It is assumed that the noise vector $\mathbf{n}$ contains  independent and identically distributed (i.i.d.) complex Gaussian elements, denoted as $\mathbf{n}\sim \mathcal{CN}(0, \sigma^{2}\mathbf{I})$.

 For the on-grid model described in (\ref{signle_snapshot_one_bit_data_model}), the likelihood function is given by\cite{gianelli2019one, shang2021weighted}:
\begin{align}  
p\left(\overline{\mathbf{y}}|\mathbf{x}\right) = \prod_{m=1}^{M} & \Phi\left(\frac{\Re\left(\overline{y}_m\right)\Re(\mathbf{a}_m^\mathrm{T} \mathbf{x})}{\sigma/\sqrt{2}}\right) 
 \Phi\left(\frac{\Im\left(\overline{y}_m\right )\Im(\mathbf{a}_m^\mathrm{T}\mathbf{x})}{\sigma/\sqrt{2}}\right),
\label{likelihood function1}
\end{align}
where $\Phi\left(\cdot\right)$ denotes the CDF of the standard normal distribution, $\overline{y}_m$ is the $m$-th element in $\overline{\mathbf{y}}$, $\mathbf{a}_m^\mathrm{T}$ is the $m$-th row vector in $\mathbf{\mathcal{A}}$. Since $\sigma > 0$ and scaling $\mathbf{x}$ by a positive constant does not affect the values of $\overline{\bm{y}}$ or the DOA estimation results,  
letting $\hat{\mathbf{x}}=\frac{\sqrt{2}}{\sigma}\mathbf{x}$, (\ref{likelihood function1}) is reformulated as:
\begin{align}  
p\left(\overline{\mathbf{y}}|\hat{\mathbf{x}}\right) = \!\prod_{m=1}^{M} & \Phi\left(\Re(\overline{y}_m)\Re(\bm{a}_m^\mathrm{T} \hat{\mathbf{x}})\right) 
\Phi\left(\Im(\overline{y}_m)\Im(\bm{a}_m^\mathrm{T} \hat{\mathbf{x}})\right).
\label{likelihood function1_new}
\end{align}
To promote sparsity in $\hat{\mathbf{x}}$, an appropriate prior PDF is required. We adopt a Laplacian-inspired prior PDF defined as \cite{5256324}: 
\begin{equation}
p\left(\hat{\mathbf{x}}\right) = \prod_{i=1}^{N}\mathrm{exp}\left(-\frac{\gamma \left|\hat{x}_i\right|^{\alpha}}{\alpha}\right), \ \ 0<\alpha\le 1,
\label{priorx}
\end{equation}
where the parameter $\gamma > 0$. As $\alpha$ approaches $0$, $p\left(\hat{\mathbf{x}}\right)$ sharply peaks at $\hat{\mathbf{x}}=0$, enforcing sparsity on the estimated  $\hat{\mathbf{x}}$. Notice that, if we choose the prior PDF as $p\left(\hat{\mathbf{x}}\right) =\frac{1}{\left|\hat{\mathbf{x}}\right|^{2}+\varepsilon}$, where $\varepsilon>0$ is a small constant, we will derive the 1bSLIM approach proposed in \cite{shang2021weighted}. Using Bayes' rule, the MAP estimator for on-grid model is given as: 
\begin{align}
\hat{\mathbf{x}}^{\star} = \mathrm{arg}\min_{\hat{\mathbf{x}}} \left\{ -\mathrm{ln} \, p\left(\overline{\mathbf{y}}|\hat{\mathbf{x}}\right) - \mathrm{ln} \, p\left(\hat{\mathbf{x}}\right) \right\}.
\label{MAP estimator on-grid}
\end{align}

Substituting (\ref{likelihood function1}) and (\ref{priorx}) into equation (\ref{MAP estimator on-grid}), we obtain the following cost function to be minimized:
\begin{align}
\mathcal{L}_{\mathrm{ongrid}} &= \sum_{m=1}^{M} \left \{ -\mathrm{ln} \, \Phi\left(\Re\left(\overline{y}_m\right )\Re(\bm{a}_m^\mathrm{T}\hat{\mathbf{x}}) \right) \right. \nonumber\\
& \ \ -\mathrm{ln} \, \Phi \left(\Im\left(\overline{y}_m\right)\Im(\bm{a}_m^\mathrm{T}\hat{\mathbf{x}}) \right)\left.+\sum_{i=1}^{N}\frac{\gamma\left|x_i\right|^{\alpha}}{\alpha} + \textrm{const}\right \}.
\label{obj_func1}
\end{align}
Since  $\mathcal{L}_{\mathrm{ongrid}}$ in (\ref{obj_func1}) is non-convex, we apply convex relaxation to make it more tractable. Specifically, using the MM principle and following the derivations given in \cite{shang2021weighted, ren2019sinusoidal}, we find the upper bound of the first two terms in (\ref{obj_func1}) as: 
\begin{align}
&\sum_{m=1}^{M}\!\left\{\!-\mathrm{ln} \, \Phi\left(\Re\left(\overline{y}_m\right )\!\Re(\bm{a}_m^\mathrm{T}\hat{\mathbf{x}})\right)\!-\!\mathrm{ln} \, \Phi \left(\Im\left(\overline{y}_m\right )\!\Im(\bm{a}_m^\mathrm{T}\hat{\mathbf{x}})\right)\!\right\} \nonumber \\
\le &  \sum_{m=1}^{M}\bigg \{\frac{1}{2} \left(\Re\left(\overline{y}_m\right) \Re(\bm{a}_m^\mathrm{T}\hat{\mathbf{x}}) \right)^{2} \!+ \frac{1}{2}\left(\Im(\overline{y}_m)\Im(\bm{a}_m^\mathrm{T}\hat{\mathbf{x}}) \right)^{2} \nonumber \\
&-\Re(v_{m}^{k})\Re(\overline{y}_m)\Re(\bm{a}_m^\mathrm{T}\hat{\mathbf{x}}) \!-\! \Im(v_{m}^{k})\Im(\overline{y}_m)\Im(\bm{a}_m^\mathrm{T}\hat{\mathbf{x}})+ c' \bigg. \bigg \},
\label{MM_approx1}
\end{align}
where $v_{m}^{k}=\left[\Re(\overline{y}_m)\Re(\tilde{v}_{m}^{k})\right] + j\left[\Im(\overline{y}_m)\Im(\tilde{v}_{m}^{k}) \right]$, $\tilde{v}_{m}^{k}=d_{m}^{k}-\mathrm{I}'(d_{m}^{k})$, $d_{m}^{k}=\Re(\overline{y}_m)\Re(\bm{a}_m^\mathrm{T}\hat{\mathbf{x}}^{k})+j\Im(\overline{y}_m)\Im(\bm{a}_m^\mathrm{T}\hat{\mathbf{x}}^{k})$, and $c'$ is a constant. Here, 
superscript $(\cdot)^{k}$ indicates variables obtained in the $k$th iteration, and function $\mathrm{I}'(x)$ is defined as: 
\begin{equation}
\mathrm{I}'(x)=-\frac{\mathrm{exp}\left(-{\Re{(x)}^{2}}/{2}\right)}{\sqrt{2\pi}\Phi(\Re{(x)})}-j\frac{\mathrm{exp}\left(-\Im{(x)}^{2}/2\right)}{\sqrt{2\pi}\Phi(\Im{(x)})}.
\end{equation}

The third term in Equation (\ref{obj_func1}) is non-convex, but can be relaxed to a convex form via smooth approximation\cite{7547360}: 
\begin{align}
\sum_{i=1}^{N}\frac{\gamma\left|x_i\right|^{\alpha}}{\alpha}\approx\frac{\gamma}{\alpha}\sum_{i=1}^{N}\left(\left|x_i\right|^{2}+\eta \right)^{\frac{\alpha}{2}},\label{smooth_approx1}
\end{align}
where $\eta>0$ is a small constant. A smaller value of $\eta$ provides a closer approximation but may reduce the robustness of the algorithm. Typically, $\eta$ is set to $10^{-6}$. 
The smoothing technique approximates the original non-smooth, non-convex objective with a differentiable surrogate. While this surrogate remains non-convex, the smoothing facilitates the application of optimization algorithms that require differentiability, such as the IRLS method\cite{6650040,7547360}.  Specifically, in the context of IRLS, smoothing allows for the formulation of weight updates and iterative schemes that can converge to stationary points of the original non-convex problem\cite{6650040}.

Substituting (\ref{MM_approx1}) and (\ref{smooth_approx1}) into (\ref{obj_func1}) yields: 
\begin{align}
\hat{\mathbf{x}}^{\star} = \mathrm{arg}\min_{\hat{\mathbf{x}}} \left  \{ \frac{1}{2}\left\|\mathbf{\mathcal{A}}\hat{\mathbf{x}} - \mathbf{v}^{k}\right\|_{2}^{2} 
+ \frac{\gamma}{\alpha}\sum_{i=1}^{N}\left(\left|\hat{x}_i\right|^{2}+\eta\right)^{\frac{\alpha}{2}} + c' \right \},
\label{MM_ongrid_obj1}
\end{align}
where $\mathbf{v}^{k} = [v_{1}^{k}, \cdots, v_{M}^{k}]^{\mathrm T}\in \mathbb{C}^{M \times 1}$.
We have the following theorem regarding the local convergence of $\hat{\mathbf{x}}^{\star}$ in SBRI framework.

\begin{theorem}[]\label{thm1}
Let $\{ (\hat{\mathbf{x}}^k, \mathbf{v}^k) \}_{k=0}^\infty$ be the sequence generated by Algorithm~\ref{alg1}, and assume the following:
\begin{enumerate}[label=(\roman*)]
    \item The measurement matrix $\mathbf{\mathcal{A}}$ is bounded;
    \item The regularizer $\sum_{i=1}^{N} (|\hat{x}_i|^2 + \eta)^{\alpha/2}$ is finite, lower semicontinuous, and satisfies $0 < \alpha \le 1$ and $\eta > 0$;
    \item The sequence $\{\gamma^k\}$ remains bounded.
\end{enumerate}
Then the joint Lyapunov function
\[
\Phi(\hat{\mathbf{x}}^k, \mathbf{v}^k) := \beta \big\| \mathbf{v}^k - \mathcal{A}\hat{\mathbf{x}}^k \big\|_2^2 + \sum_{i=1}^M H(v_i^k) + \frac{\gamma^k}{\alpha} \sum_{i=1}^N (|\hat{x}_i^k|^2 + \eta)^{\alpha/2}
\]
with $\beta \ge 0$ is non-increasing and bounded. Consequently, the sequence $\{(\hat{\mathbf{x}}^k, \mathbf{v}^k)\}$ is bounded and admits at least one convergent subsequence.
Moreover, the limit point $(\hat{\mathbf{x}}^{\dagger}, \mathbf{v}^{\dagger})$ of any convergent subsequence is a stationary point of the alternating update system.
\end{theorem}

\begin{corollary}[]\label{cor1}
Suppose that the 1-bit quantized observations $\overline{\mathbf{y}}$ are generated from a true signal $\hat{\mathbf{x}}^{\diamondsuit}$ via
\[
\overline{\mathbf{y}} = \operatorname{csgn}(\mathbf{\mathcal{A}} \hat{\mathbf{x}}^{\diamondsuit} + \mathbf{n}),
\]
where $\hat{\mathbf{x}}^{\diamondsuit}$ is an $s$-sparse vector and $\bf{n}$ is additive noise. Assume that the measurement matrix $\mathbf{\mathcal{A}}$ satisfies the Restricted Isometry Property (RIP) of order $2s$ with constant $\delta_{2s} < 1$, and that the smoothing parameter $\eta > 0$. Let $\{\hat{\mathbf{x}}^k\}$ be the sequence produced by Algorithm~\ref{alg1} with $0 < \alpha \le 1$.
Then any convergent subsequence of $\{\hat{\mathbf{x}}^k\}$ admits a limit $\hat{\mathbf{x}}^{\dagger}$ that satisfies the recovery bound
\[
\left\lVert \hat{\mathbf{x}}^{\dagger} - \hat{\mathbf{x}}^{\diamondsuit} \right\rVert_2
\le C_1 + C_2 \cdot \sigma_s(\hat{\mathbf{x}}^{\dagger})_2,
\]
where $\sigma_s(\hat{\mathbf{x}}^{\dagger})_2 := \inf_{\|\mathbf{z}\|_0 \le s} \| \hat{\mathbf{x}}^{\dagger} - \mathbf{z} \|_2$ denotes the best $s$-sparse approximation error, and $C_1, C_2 > 0$ are constants depending only on $\delta_{2s}$ and the initial point $\hat{\mathbf{x}}^0$.
\end{corollary}

The proof of Theorem~\ref{thm1} and Corollary~\ref{cor1} is in the Appendix \ref{Appendix_A}.
Note that Equation (\ref{MM_ongrid_obj1}) represents a classic regularized least squares problem, and $\gamma$ is a regularization parameter. Therefore, we  utilize the IRLS  method\cite{daubechies2010iteratively}, leveraging the estimates obtained from the $k$-th iteration step, to compute $\hat {\mathbf{v}}^k$ and $\hat {\mathbf{x}}^{k+1}$ as:
\begin{align}
\mathbf{v}^{k} & \!\! = [\Re(\overline{\mathbf{y}})\odot\Re(\mathbf{D}^k \!\!-\mathrm{I}'(\mathbf{D}^{k}))]+j[\Im(\overline{\mathbf{y}})\odot\Im(\mathbf{D}^k \!\!-\mathrm{I}'(\mathbf{D}^{k}))],  
\label{step_v_update}
\end{align}
\begin{align}
\hat{\mathbf{x}}^{k+1} & = \left(\mathcal{A}^{\mathrm H}\mathcal{A} + \gamma^{k}\bm{\Lambda}(\hat{\mathbf{x}}^{k})\right)^{-1}\mathcal{A}^{\mathrm H}\mathbf{v}^{k},  \label{step_x_update}
\end{align}
\noindent where 
$\mathbf{D}^{k} = \left[\Re(\overline{\mathbf{y}})\odot\Re(\mathbf{\mathcal{A}}\hat{\mathbf{x}}^k)\right]+j\left[\Im(\overline{\mathbf{y}})\odot\Im(\mathbf{\mathcal{A}}\hat{\mathbf{x}}^k)\right],$ and
\begin{align}
\bm{\Lambda}(\hat{\mathbf{x}}^{k})=\mathrm{diag}\left( \left(|\hat{x}_1^{k}|^{2}+\eta \right)^{\frac{\alpha}{2}-1},\cdots, \left(|\hat{x}_N^{k}|^{2}+\eta \right)^{\frac{\alpha}{2}-1}\right). 
\end{align}
In addition, updating the regularization parameter $\gamma$ as
\begin{equation}
    \gamma^{k+1} = \gamma^{0}\|\hat{\mathbf{x}}^{k+1}\|_{2}
    \label{lambda_update}
\end{equation}
simplifies its selection and improves the robustness of the iterative process. In general, $\gamma^{0}$ is initialized to 1. 
The proposed SBRI approach for one-bit on-grid DOA estimation is summarized in Algorithm \ref{alg1}. 

\begin{algorithm}
\caption{SBRI for One-Bit On-Grid DOA}
{\begin{algorithmic}[1]
\State \textbf{Input:} $\overline{\mathbf{y}}$, 
$\mathbf{\mathcal{A}}$,
$T_{\max}$, $\alpha$, $\eta$ $\gamma^{0}$, error bound $\epsilon_{0}$

\State \textbf{Output:} $\hat{\mathbf{x}}^{\star}$

\State Initialize $\hat{\mathbf{x}}^{0} = \frac{\mathcal{A}^H \overline{\mathbf{y}}}{\|\mathcal{A}^H \overline{\mathbf{y}}\|_2}$
\For {$k = 0$ to $T_{\max}-1$}
    \State Compute $\mathbf v^{k}$ via \eqref{step_v_update}:
    $\mathbf v^{k} \gets [\Re(\overline{\mathbf{y}})\odot\Re(\mathbf{D}^k-\mathrm{I}'(\mathbf{D}^{k}))]+j[\Im(\overline{\mathbf{y}})\odot\Im(\mathbf{D}^k-\mathrm{I}'(\mathbf{D}^{k}))]$
    \State 
    Update $\hat{\mathbf{x}}^{k+1}$ via \eqref{step_x_update}:
    $\hat{\mathbf{x}}^{k+1} \gets
      (\mathcal{A}^{\mathrm H}\mathcal{A} + \gamma^k\Lambda(\hat{\mathbf{x}}^k))^{-1}
       \mathcal{A}^{\mathrm H}\mathbf{v}^k$
    \State $\gamma^{k+1} \gets \gamma^{0}\|\hat{\mathbf{x}}^{k+1}\|_{2}$
    \If{convergence($\hat{\mathbf{x}}^k,\hat{\mathbf{x}}^{k+1},\epsilon_0$)} \textbf{break} 
    \EndIf
\EndFor
\State \textbf{return} $\hat{\mathbf{x}}^{\star} \gets \hat{\mathbf{x}}^{k+1}$
\end{algorithmic}
}
\label{alg1}
\end{algorithm}

\vspace{-0.5em}
\subsection{One-Bit Off-Grid DOA Estimation via SBRI Framework}

For the off-grid model given in (\ref{one_bit_off_grid_single_snpashot_DOA_model}), the likelihood function is obtained by replacing $\bm{a}^{\T}_m$ in (\ref{likelihood function1}) by $\bm{c}_m^{\T} (\bm{\beta})$, 
where $\bm{c}_m^\mathrm{T} \left(\bm{\beta}\right)$ is the $m$-th row of $\mathbf{C}(\bm{\beta})$ representing the modified steering vector incorporating the off-grid gaps $\bm{\beta}$. Similar to the derivation in (\ref{likelihood function1_new}), by letting $\hat{\mathbf{x}}=\frac{\sqrt{2}}{\sigma}\mathbf{x}$, the likelihood function is  as: 
\begin{align}  
p\left(\overline{\mathbf{y}}|\hat{\mathbf{x}};\bm{\beta}\right) = \!\prod_{m=1}^{M} & \Phi\left(\Re(\overline{y}_m)\Re\left(\bm{c}_m^\mathrm{T} (\bm{\beta})\hat{\mathbf{x}}\right)\right) \nonumber\\
&\times 
\Phi\left(\Im(\overline{y}_m)\Im\left(\bm{c}_m^\mathrm{T} (\bm{\beta})\hat{\mathbf{x}}\right)\right).
\label{likelihood function2}
\end{align}
The off-grid gaps $\bm{\beta}$ follow a uniform distribution, $p\left(\bm{\beta}\right) \sim U\left(-\frac{r}{2}, \frac{r}{2}\right)$\cite{yang2012off}, independent of $\hat {\mathbf {x}}$, where $r$ denotes the size of the grid interval.
We adopt a uniform prior for the off-grid gaps \(\bm{\beta}\) for three reasons. First, with no deterministic knowledge of the true DOAs, the uniform distribution is the non-informative (maximum-entropy) choice, assigning equal likelihood across each grid cell. Second, this matches the assumptions of leading off-grid sparse Bayesian learning methods, e.g., OGSBI \cite{yang2012off}, off-grid SBL-RVM \cite{7859460}, and OGIR \cite{10024794}, thus ensuring a fair comparison in Section~V. Finally, the uniform prior admits simple closed-form MAP updates and directly yields the efficient update rules in Equations~\eqref{step_w_update}--\eqref{update_grid_gaps} and Equations~\eqref{x_update_offgrid_modify}--\eqref{beta_update_offgrid_modify}, whereas alternatives (e.g., Gaussian or Beta priors) would introduce extra hyperparameters and necessitate numerical integration. 
Using Bayes' rule, the MAP estimator for the off-grid model is given by: 
\begin{align}
\left\{\hat{\mathbf{x}}^{\star}, \bm{\beta}^{\star}\right\} = \mathrm{arg}\min_{\hat{\mathbf{x}}, \bm{\beta}} \bigg \{-\mathrm{ln} \, p\left(\mathbf{y}|\hat{\mathbf{x}};\bm{\beta}\right) - \mathrm{ln} \, p\left(\hat{ \mathbf{x}}\right) - \mathrm{ln} \, p\left(\bm{\beta}\right) \bigg \}.
\label{MAP estimator off-grid}
\end{align}
Substituting (\ref{likelihood function2}) and (\ref{priorx}) into   (\ref{MAP estimator off-grid}), we obtain the following cost function to minimize:
\begin{align}
\mathcal{L}_\mathrm{offgrid} &= \sum_{m=1}^{M} \left \{ -\mathrm{ln} \, \Phi\left(\Re\left(\overline{y}_m\right )\Re (\bm{c}_m^\mathrm{T}(\bm{\beta})\hat{\mathbf{x}} ) \right) \right. \nonumber\\ 
& \ -\mathrm{ln} \, \Phi \left(\Im\left(\overline{y}_m\right)\Im(\bm{c}_m^\mathrm{T}(\bm{\beta})\hat{\mathbf{x}} ) \right)\left.\!+\!\sum_{i=1}^{N}\frac{\gamma\left|x_i\right|^{\alpha}}{\alpha} \!+\! \textrm{const}\right\}.
\label{obj_func2}
\end{align}
Following the same approach as in (\ref{MM_approx1})--(\ref{smooth_approx1}), we derive the new minimization problem, defined as:
\begin{align}
\left\{\hat{\mathbf{x}}^{\star}, \bm{\beta}^{\star}\right\} = \mathrm{arg}\min_{\hat{x}, \beta} & \  \bigg \{ \frac{1}{2}\left\|\mathbf{C}(\bm{\beta})\hat{\mathbf{x}} - \mathbf{w}^{k}\right\|_{2}^{2} \nonumber \\ & + \frac{\gamma}{\alpha}\sum_{i=1}^{N}\left(\left|\hat{x}_i\right|^{2}+\eta\right)^{\frac{\alpha}{2}} + c'  \bigg \},
\label{obj_offgrid_original}
\end{align}
where $\mathbf{w}^{k} = [w_{1}^{k}, \cdots, w_{M}^{k}]^{\mathrm T}$, $w_{m}^{k}=\left[\Re(\overline{y}_m)\Re(\tilde{w}_{m}^{k})\right] + j\left[\Im(\overline{y}_m)\Im(\tilde{w}_{m}^{k}) \right]$, $\tilde{w}_{m}^{k}=e_{m}^{k}-\mathrm{I}'(e_{m}^{k})$, $e_{m}^{k}=\Re(\overline{y}_m)\Re(\bm{c}_m^\mathrm{T}(\bm{\beta}^{k})\hat{\mathbf{x}}^{k})+j\Im(\overline{y}_m)\Im(\bm{c}_m^\mathrm{T}(\bm{\beta}^{k})\hat{\mathbf{x}}^{k})$.
By employing an alternating iterative approach, we first compute $\mathbf{w}^k$ and then update the spectral coefficients $\hat{\mathbf{x}}$ and the grid gaps $\bm{\beta}$ as follows:
\begin{align}
\mathbf{w}^{k} & = [\Re(\overline{\mathbf{y}})\odot\Re(\mathbf{E}^k-\mathrm{I}'(\mathbf{E}^{k}))]+j[\Im(\overline{\mathbf{y}})\odot\Im(\mathbf{E}^k-\mathrm{I}'(\mathbf{E}^{k}))],  
\label{step_w_update}
\end{align}
\begin{align}
\hat{\mathbf{x}}^{k+1}  = & \left[\mathbf{C}^{\mathrm H}(\bm{\beta}^{k})\mathbf{C}(\bm{\beta}^{k}) + \gamma^{k}\bm{\Lambda}(\hat{\mathbf{x}}^{k})\right]^{-1}\mathbf{C}^{\mathrm H}(\bm{\beta}^{k})\mathbf{w}^{k},  \label{update_x_step1_offgrid}
\\
\bm{\beta}^{k+1}  = & 
\ \left [\Re \left((\mathcal{B}^{\mathrm{H}}\mathcal{B})^{*}\hat{\mathbf{x}}^{k+1}(\hat{\mathbf{x}}^{k+1})^{\mathrm H}\right)\right]^{-1}  \nonumber\\
&\times\Re\left(\mathrm{diag}\left((\hat{\mathbf{x}}^{k+1})^{*}\right)\mathcal{B}^{\mathrm H}\left[\mathbf{w}^{k} - \mathcal{A}\hat{\mathbf{x}}^{k+1}\right]\right),\label{update_grid_gaps}
\end{align}
where \ $\mathbf{E}^{k} =[\Re(\overline{\mathbf{y}})\odot\Re(\mathbf{C}(\bm{\beta}^k)\hat{\mathbf{x}}^k)]+j[\Im(\overline{\mathbf{y}})\odot\Im(\mathbf{C}(\bm{\beta}^k)\hat{\mathbf{x}}^k)].$

The derivations for (\ref{update_x_step1_offgrid}) and (\ref{update_grid_gaps}) are as follows. First, we compute the first-order derivative of the objective function in Equation (\ref{obj_offgrid_original}) with respect to $\hat{\mathbf{x}}$ as:
\begin{align}
\frac{\partial{\mathcal{L}}\left(\hat{\mathbf{x}}, \bm{\beta}^k\right)}{\partial \hat{\mathbf{x}}} =&\mathbf{C}^{\mathrm H}(\bm{\beta}^{k})\left[\mathbf{C}(\bm{\beta}^{k})\hat{\mathbf{x}} - \mathbf{w}^{k}\right] \nonumber + \gamma^{k}\bm{\Lambda}(\hat{\mathbf{x}})\hat{\mathbf{x}}.
\end{align}
Setting the derivative to zero, we obtain:
\begin{align}
\left[\mathbf{C}^{\mathrm H}(\bm{\beta}^{k})\mathbf{C}(\bm{\beta}^{k})+\gamma^{k}\bm{\Lambda}(\hat{\mathbf{x}})\right]\hat{\mathbf{x}} =\mathbf{C}^{\mathrm H}(\bm{\beta}^{k})\mathbf{w}^{k}.
\label{first_order_equ_offgrid_original}
\end{align}
Since $\bm{\Lambda}(\hat{\mathbf{x}})$ is a nonlinear function of $\hat{\mathbf{x}}$, Equation (\ref{first_order_equ_offgrid_original}) is inherently nonlinear, making it challenging to solve directly. However, an iterative approximation method can be utilized to efficiently estimate its solution. In this approach, at each iteration, Equation (\ref{first_order_equ_offgrid_original}) is approximated as a following linear problem:
\begin{align}
\left[\mathbf{C}^{\mathrm H}(\bm{\beta}^{k})\mathbf{C}(\bm{\beta}^{k})+\gamma^{k}\bm{\Lambda}(\hat{\mathbf{x}}^k)\right ]\hat{\mathbf{x}} =\mathbf{C}^{\mathrm H}(\bm{\beta}^{k})\mathbf{w}^{k}.
\label{first_order_equ_offgrid_original_approx}
\end{align}
The solution of (\ref{first_order_equ_offgrid_original_approx}) yields Equation (\ref{update_x_step1_offgrid}).
With the estimate $\hat{\mathbf{x}}^{k+1}$ obtained from Equation (\ref{update_x_step1_offgrid}), we can further estimate $\bm{\beta}^{k+1}$. Rearrange the objective function as: 
\begin{align} \mathcal{L}\left(\hat{\mathbf{x}}, \bm{\beta}^k\right) =& \frac{1}{2}\biggl(\bm{\beta}^{\mathrm T}\mathrm{diag}(\hat{\mathbf{x}}^{k+1})^{*}\mathbf{\mathcal{B}}^{\mathrm{H}} \mathbf{\mathcal{B}}\mathrm{diag}(\hat{\mathbf{x}}^{k+1} )\bm{\beta} \biggr)\nonumber \\  &+\bm{\beta}^{\mathrm T}\mathrm{diag}(\hat{\mathbf{x}}^{k+1})^{*}\mathbf{\mathcal{B}}^{\mathrm{H}}\left[\mathbf{\mathcal{A}}\hat{\mathbf{x}}^{k+1}-\mathbf{w}^{k} \right] \nonumber \\
    &+
\left((\hat{\mathbf{x}}^{k+1})^{\mathrm{H}}\mathbf{\mathcal{A}}^{\mathrm{H}} - (\mathbf{w}^{k})^{\mathrm{H}} \right)\mathbf{\mathcal{B}}\mathrm{diag}(\hat{\mathbf{x}}^{k+1})\bm{\beta}\biggr) + C', 
    \label{obj_beta_related}
\end{align}
where $C'$ represents the sum of all terms that are not related to $\bm{\beta}$. Taking the first-order derivative of ($\ref{obj_beta_related}$) with respect to $\bm{\beta}$ and setting it to zero, we have:
\begin{align}
&\Re\left((\mathcal{B}^{\mathrm{H}}\mathcal{B})^{*}\hat{\mathbf{x}}^{k+1}(\hat{\mathbf{x}}^{k+1})^{\mathrm H}\right)\bm{\beta} \nonumber\\
= & \Re\left(\mathrm{diag}\left((\hat{\mathbf{x}}^{k+1})^{*}\right)\mathcal{B}^{\mathrm H}\left[\mathbf{w}^{k} - \mathcal{A}\hat{\mathbf{x}}^{k+1}\right]\right).
\end{align}
Thus, we have the estimate of $\bm{\beta}^{k+1}$, given in Equation (\ref{update_grid_gaps}).

The proposed SBRI for the one-bit off-grid model is summarized in Algorithm \ref{alg2}. 

\begin{algorithm}[h]
\caption{SBRI for One-Bit Off-Grid DOA.}
{\begin{algorithmic}[1]
\State \textbf{Input:} $\overline{\mathbf{y}}$, 
$\mathbf{\mathcal{A}}$,
$\mathbf{\mathcal{B}}$,
$T_{\max}$, $\alpha$, $\eta$, $\gamma^{0}$, error bound $\epsilon_{0}$

\State \textbf{Output:} $\hat{\mathbf{x}}^{\star}$, $\bm{\beta}^{\star}$

\State Initialize $\hat{\mathbf{x}}^{0} = \frac{\mathcal{A}^{\mathrm H} \overline{\mathbf{y}}}{\|\mathcal{A}^{\mathrm H} \overline{\mathbf{y}}\|_2}$, $\bm{\beta}^{0} = \bm{0}$ \label{init_step_alg2}
\For{$k = 0$ to $T_{\max}-1$}
    \State Compute $\mathbf{w}^k$ via (\ref{step_w_update}): $\mathbf{w}^k \gets [\Re(\overline{\mathbf{y}})\odot\Re(\mathbf{E}^k-\mathrm{I}'(\mathbf{E}^{k}))]+j[\Im(\overline{\mathbf{y}})\odot\Im(\mathbf{E}^k-\mathrm{I}'(\mathbf{E}^{k}))]$.
    \State Update $\hat{\mathbf{x}}^{k+1}$ via  (\ref{update_x_step1_offgrid}):
    $\hat{\mathbf{x}}^{k+1} \gets
\bigl(\mathbf{C}^{\mathrm H}(\bm\beta^k)\mathbf{C}(\bm\beta^k)+\gamma^k\bm{\Lambda}(\hat{\mathbf{x}}^k)\bigr)^{-1}
       \mathbf{C}^{\mathrm H}(\bm\beta^k)\mathbf{w}^k$
    \State  
    $\bm\beta^{k+1}\gets \text{gridUpdate}_{\text{SBRI}}(\hat{\mathbf{x}}^{k+1},\mathcal{A},\mathcal{B},\mathbf{w}^k)$, $\text{gridUpdate}_{\text{SBRI}}(\cdot)$ is defined in (\ref{update_grid_gaps}).
    \label{grid_gaps_update}
    \State  $\gamma^{k+1}\gets \gamma^0\|\hat{\mathbf{x}}^{k+1}\|_2$
    \If{convergence$(\hat{\mathbf{x}},\bm\beta,\epsilon_0)$} \textbf{break} \EndIf
\EndFor
\State \textbf{return} $\hat{\mathbf{x}}^{\star} \gets \hat{\mathbf{x}}^{k+1}$, $\bm{\beta}^{\star} \gets \bm{\beta}^{k+1}$
\end{algorithmic}}
\label{alg2}
\end{algorithm}

\vspace{-5mm}
\subsection{One-Bit DOA Estimation via Augmented Framework: SBRI-X}
\label{the augmented framework}

\begin{figure}[h]
\centering
\includegraphics[width=0.3\textwidth]{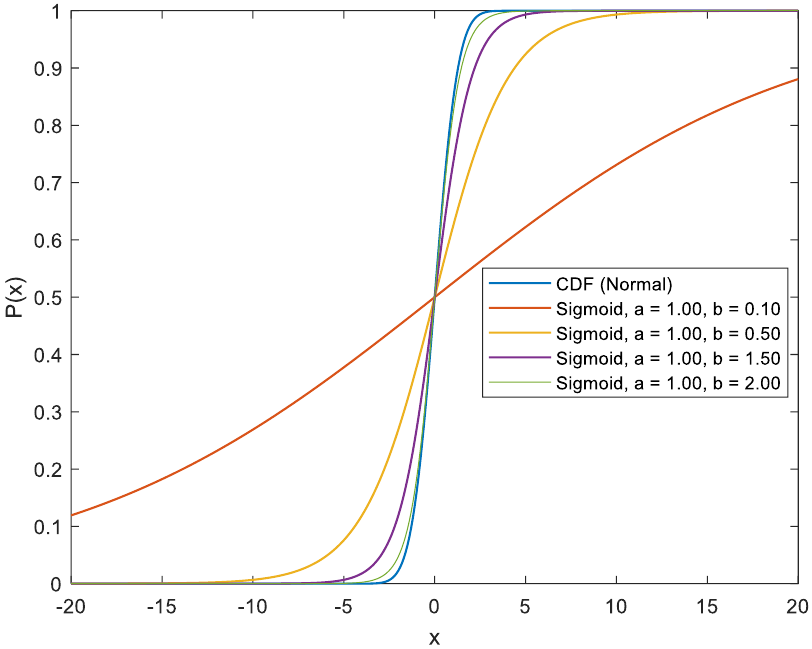}\\
\caption{Different types of likelihood function.}
\label{example_different_likelihood}
\vspace{-0.5em}
\end{figure}

We treat the one-bit sensing problem as a binary classification problem in which the likelihood follows the Bernoulli distribution with a sigmoid link function.
Specifically, the Bernoulli-type likelihood model of $\overline{\mathbf{y}}$, given the input $\mathbf{y} = \mathbf{A}(\bm{\theta})\mathbf{x} + \mathbf{n}$, is expressed as:
\begin{equation}
p\left(\overline{\mathbf{y}} \mid \mathbf{y}\right)  = \prod_{m=1}^{M} \left[\mathrm{sig}\left(y_{m} \right) \right]^{\frac{1+\overline{y}_{m}}{2} }\left[1-\mathrm{sig}\left(y_{m}\right)\right]^{\frac{1-\overline{y}_{m}}{2} }. \label{binary_likelihood}
\end{equation} 
It is straightforward to verify that 
\begin{equation}
\left[\mathrm{sig}\left(y_{m}\right) \right]^{\frac{1+\overline{y}_{m}}{2} }\left[1-\mathrm{sig}\left(y_{m}\right)\right]^{\frac{1-\overline{y}_{m}}{2}}=\mathrm{sig}\left(y_{m}\overline{y}_{m}\right),
\end{equation}
where $\mathrm{sig}\left(x\right)$ is the sigmoid link function, defined as
\begin{equation}
\mathrm{sig}\left(x\right) \triangleq \frac{1}{1+\mathrm{a}\ \mathrm{exp}(-\mathrm{b}x)},
\end{equation}
and parameters $\mathrm{a}$ and $\mathrm{b}$ control the shape of the sigmoid function. As illustrated in Figure \ref{example_different_likelihood}, the parameter $\mathrm{b}$ governs the slope of the likelihood function in the vicinity of the zero point. A larger value of $\mathrm{b}$ results in a steeper slope, making the sigmoid function more closely approximate the CDF.
The sigmoid likelihood function is inherently smooth and differentiable, thus is well-suited for optimization tasks.  Compared to the CDF likelihood, the sigmoid function offers greater computational efficiency and seamless integration into neural networks for training purposes. By tuning the parameters of the sigmoid function, it can effectively model the impact of noise, thereby enhancing performance, particularly in low SNR conditions. Furthermore, the sigmoid likelihood function is highly versatile and can be readily extended to address more complex scenarios, such as non-Gaussian noise or colored noise, making it a robust choice for a wide range of applications.

Denote $\tilde{\mathbf{x}}=\frac{\mathbf{x}}{\sigma}$, and notice 
\begin{align}
P\left(\overline{\mathbf{y}}=1\right) =  P\left(\mathbf{\mathcal{A}}\mathbf{x} + \mathbf{n}\ge 
 0\right) = P\left(\mathbf{\mathcal{A}} \tilde {\mathbf{x}} + \bm{\epsilon}\ge 
 0\right),
\end{align}
where $\bm{\epsilon}$ is the normalized noise term that follows a standard normal distribution. 
Similarly,
\begin{align}
P\left(\overline{\mathbf{y}}=-1\right)
 =  P\left(\mathbf{\mathcal{A}} \tilde{\mathbf{x}} + \bm{\epsilon} < 
 0\right).
\end{align}
Thus, the posterior probability is constructed as:
\begin{equation}
p\left(\tilde{\mathbf{x}} \mid \overline{\mathbf{y}},\bm{\epsilon} \right) \propto p\left(\overline{\mathbf{y}} \mid \tilde{\mathbf{x}},\bm{\epsilon} \right)p\left(\tilde{\mathbf{x}}\right).
\end{equation}
We can estimate $\tilde{\mathbf{x}}$ and $\bm{\epsilon}$ via the MAP approach as:
\begin{equation}
 \left(\tilde{\mathbf{x}}^{\star}, \bm{\epsilon}^{\star}\right) = \mathrm{arg}\max_{\tilde{\mathbf{x}}, \bm{\epsilon}} p\left(\overline{\mathbf{y}} \mid \tilde{\mathbf{x}},\bm{\epsilon} \right)p\left(\tilde{\mathbf{x}}\right).
\label{post_prob}
\end{equation}
Following the same approach as in Section \ref{original_framework}, taking the negative logarithm of Equation (\ref{post_prob}) yields the following non-convex objective function: 
\begin{align}
\mathcal{L}_{ongrid}^{\diamond}\left(\tilde{\mathbf{x}}, \bm{\epsilon}\right) =& -\ln_{}{p\left(\overline{\mathbf{y}} \mid \tilde{\mathbf{x}},\bm{\epsilon}\right)} -\ln_{}{p\left(\tilde{\mathbf{x}}\right)} \nonumber\\
= &\sum_{i=1}^{N}\ln_{}{\left(1+\mathrm{a}\exp\left(-\mathrm{b}\overline{y}_i\left[\mathbf{a}_{i}^{\mathrm{T}}\tilde{\mathbf{x}}+\epsilon_i\right]\right)\right)} \nonumber \ \\
&+\sum_{i=1}^{N}\ln_{}{\left(1+\mathrm{a}\exp\left(-\mathrm{b}\overline{y}_i\left[\mathbf{a}_{i}^{\mathrm{T}}\tilde{\mathbf{x}}+\epsilon_i\right]\right)\right)} \nonumber\\
&+\sum_{i=1}^{N}\frac{\gamma
\left|x_i\right|^{\alpha}}{\alpha}.
    \label{obj_fun}
\end{align}
The same MM approach can be utilized  to derive a majorizing function for $\mathcal{L}_{ongrid}^{\diamond}\left(\tilde{\mathbf{x}}, \bm{\epsilon}\right)$. This leads to the derivation of a new closed-form update formula for each MM iteration. 

\subsubsection{Majorizing Function for 
$\mathcal{L}_{ongrid}^{\diamond}\left(\tilde{\mathbf{x}}, \bm{\epsilon}\right)$} \ 

Let $f\left(s\right) = \ln_{}{\left(1+\mathrm{a}\exp\left(-\mathrm{b}s\right)\right)}$. With the second-order Taylor expansion of $f\left(s\right)$,  we have:
\begin{equation}
    f\left(s\right) \approx  f\left(s^{k}\right) + f'\left(s^{k}\right)\left(s-s^{k}\right) +\frac{f''\left(s^{k}\right)}{2}\left(s-s^{k}\right)^{2},
\end{equation}
where $s^{k}$ is assumed to be available from the $k$-th iteration. 
The second-order derivative  $f''(s^{k})$ is upper bounded by:
\begin{equation}
    f''\left(s^{k}\right) = \frac{\mathrm{a}\mathrm{b}^2}{\exp(\mathrm{b}s^k)+\mathrm{a}^2\exp(-\mathrm{b}s^{k})+2\mathrm{a}} \le \frac{\mathrm{a}\mathrm{b}^2}{(\mathrm{a} + 1)^2}.   
\end{equation}
Thus, $f\left(s\right)$ is upper bounded by:
\begin{equation}
    f\left(s\right) \le \frac{\mathrm{a}\mathrm{b}^2}{2(\mathrm{a}+1)^2}\left(s+\frac{(\mathrm{a}+1)^2}{\mathrm{a}\mathrm{b}^2}f'(s^{k})-s^{k} \right)^{2}+\mathrm{const} .\label{secondorder_ub}     
\end{equation}
By substituting (\ref{secondorder_ub}) into (\ref{obj_fun}) and applying smooth approximation on the second term in (\ref{obj_fun}),
we obtain the majorizing function $\bar{\mathcal{L}}_{ongrid}^{\diamond}\left(\tilde{\mathbf{x}}, \bm{\epsilon}\right)$ of $\mathcal{L}_{ongrid}^{\diamond}\left(\tilde{\mathbf{x}}, \bm{\epsilon}\right)$:
\begin{align}
\mathcal{L}_{ongrid}^{\diamond}\left(\tilde{\mathbf{x}}, \bm{\epsilon}\right) \le &\bar{\mathcal{L}}_{ongrid}^{\diamond}\left(\tilde{\mathbf{x}}, \bm{\epsilon}\right) \nonumber\\
=& \frac{\mathrm{a}\mathrm{b}^2}{2(\mathrm{a}+1)^2}\left\|\mathbf{\mathcal{A}}\tilde{\mathbf{x}}-\mathbf{\mathcal{A}}\tilde{\mathbf{x}}^{k} + \bm{\epsilon} - \bm{\epsilon}^k -g(\tilde{\mathbf{x}}^{k}, \bm{\epsilon}^k)\right\|^{2} \nonumber\\ 
&+\frac{\gamma}{\alpha}\sum_{i=1}^{N}\left(\left|\tilde{x}_i\right|^{2}+\eta\right)^{\frac{\alpha}{2}} + \mathrm{const},
\label{obj_MM_augmented}
\end{align}
where 
\begin{align} g(\tilde{\mathbf{x}}^{k},\bm{\epsilon}^{k}) =&\frac{(\mathrm{a}+1)^2\Re({\overline{\mathbf{y}}})} {\mathrm{b}\exp\left(\mathrm{b} \Re \left({\overline{\mathbf{y}}} \right)\odot\Re \left(\mathbf{\mathcal{A}}\tilde{\mathbf{x}}^k+\bm{\epsilon}^k\right) \right)+\mathrm{a}\mathrm{b}} \nonumber \\ 
    &+ j\frac{(\mathrm{a}+1)^2\Im({\overline{\mathbf{y}}})} {\mathrm{b}\exp\left(\mathrm{b}\Im({\overline{\mathbf{y}}})\odot \Im \left(\mathbf{\mathcal{A}}\tilde{\mathbf{x}}^k+\bm{\epsilon}^k \right) \right)+\mathrm{a}\mathrm{b}}.
\end{align}
We have the following theorem regarding the convergence of $\tilde{\mathbf{x}}^{\star}$ in the SBRI-X framework.
\begin{theorem}[]\label{thm2}
Suppose that $\tilde{\mathbf{x}}^{\diamondsuit}$ is an $s$-sparse vector.  Assume that $\mathbf{\mathcal{A}}$ obeys the RIP\cite{foucart2012sparse} of order
$2s$ with constant $\delta_{2s}<1$ and a smoothing parameter
$\eta > 0$.  Let
$\{\tilde{\mathbf{x}}^{k}\}$ be the sequence generated by Algorithm 2 with
$0<\alpha\le1$.  Then $\{\tilde{\mathbf{x}}^{(k)}\}$ has at least one convergent
subsequence.  When $\eta>0$, the limit
$\tilde{\mathbf{x}}^{\dagger}$ of any convergent subsequence is
a limit point of the problem in Equation (\ref{obj_MM_augmented}) and satisfies
\begin{equation}\label{theorem1_ineq1}
  \left\lVert \tilde{\mathbf{x}}^{\dagger}-\tilde{\mathbf{x}}^{o}\right\rVert_{2}
  \;\le\;
  C_{1}'\sqrt{\frac{2(\mathrm{a}+1)}{\mathrm{a}\mathrm{b}^2}}\;+\;C_{2}'\,
  \sigma_{s}\!\bigl(\tilde{\mathbf{x}}^{{\dagger}}\bigr)_{2},
\end{equation}
where $C_{1}',C_{2}'>0$ depend only on $\delta_{2s}$ and  initial point $\hat{\mathbf{x}}^{0}$.
\end{theorem}

The proof of Theorem \ref{thm2} is included in Appendix \ref{Appendix_B}.

\subsubsection{Updating Formula} \ 

The alternative minimization procedure is utilized to update $\tilde{\mathbf{x}}$ and  $\bm{\epsilon}$. Specifically, $\tilde{\mathbf{x}}$ and $\bm{\epsilon}$ are updated iteratively by solving the following subproblems:
\begin{align}
{\tilde{\mathbf{x}}^{k+1}} &= \mathrm{arg}\min_{\tilde{\mathbf{x}}}\bar{\mathcal{L}}_{ongrid}^{\diamond}\left(\tilde{\mathbf{x}}, \bm{\epsilon}^{k}\right),
    \label{subopt1L_up}\\
    \bm{\epsilon}^{k+1} &= \mathrm{arg}\min_{\bm{\epsilon}}\bar{\mathcal{L}}_{ongrid}^{\diamond}\left(\tilde{\mathbf{x}}^{k+1}, \bm{\epsilon}\right).
    \label{subopt2L_up}
\end{align}
For subproblem (\ref{subopt1L_up}), following the same approach as in (\ref{step_x_update}) or (\ref{update_x_step1_offgrid}), we derive the estimate of $\tilde{\mathbf{x}}$:
\begin{align}
\tilde{\mathbf{x}}^{k+1} \!=\!\left[\mathbf{\mathcal{A}}^{\mathrm H}\mathbf{\mathcal{A}}+\frac{\gamma^{k}(\mathrm{a}+1)^2}{\mathrm{a}\mathrm{b}^2}\bm{\Lambda}(\tilde{\mathbf{x}}^{k})\right]^{-1}\!\!\mathbf{\mathcal{A}}^{\mathrm H}\left[\mathbf{\mathcal{A}}\tilde{\mathbf{x}}^{k}+g(\tilde{\mathbf{x}}^{k},\bm{\epsilon}^k)\right],
    \label{x_update_modify}
\end{align}
where 
\begin{equation}
    \bm{\Lambda} \left(\tilde{\mathbf{x}}^{k} \right)=\mathrm{diag}\left( \left(\left|\tilde{x}_1^k\right|^{2}+\eta \right)^{\frac{\alpha}{2}-1},\cdots, \left(\left|\tilde{x}_N^k\right|^{2}+\eta \right)^{\frac{\alpha}{2}-1}\right). \nonumber
\end{equation}
We then alternatively update $\bm{\epsilon}^{k+1}$ as:
\begin{equation}
    \bm{\epsilon}^{k+1} = \bm{\epsilon}^{k}-\mathbf{\mathcal{A}}\left[\tilde{\mathbf{x}}^{k+1} - \tilde{\mathbf{x}}^{k} \right]+g(\tilde{\mathbf{x}}^{k},\bm{\epsilon}^k).
    \label{epsi_update_modify}
\end{equation}
The augmented framework, referred to as sparse Bayesian reweighted iterative X (SBRI-X) for one-bit on-grid DOA estimation, is summarized in Algorithm \ref{alg3}.
\begin{algorithm}
\caption{{SBRI-X for One-Bit On-Grid DOA.}}
{{\begin{algorithmic}[1]
\State \textbf{Input:} $\overline{\mathbf{y}}$, 
$\mathbf{\mathcal{A}}$,
$T_{\max}$, $\alpha$, $\mathrm{a}$, $\mathrm{b}$, $\eta$ $\gamma^{0}$, error bound $\epsilon_{0}$
\State \textbf{Output:} $\tilde{\mathbf{x}}^{\star}$
\State Initialize $\tilde{\mathbf{x}}^{0} = \frac{\mathcal{A}^H \overline{\mathbf{y}}}{\|\mathcal{A}^H \overline{\mathbf{y}}\|_2}$
\For{$k = 0$ to $T_{\max}-1$}
    \State $\tilde{\mathbf{x}}^{k+1}\!\gets\!(\mathcal{A}^H\mathcal{A}+\tfrac{\gamma^k(a+1)^2}{ab^2}\bm{\Lambda}(\tilde{\mathbf{x}}^{k}))\!^{-1}\!\mathcal{A}^{\mathrm H}[\mathcal{A}\tilde{\mathbf{x}}^k+g(\tilde{\mathbf{x}}^{k},\bm{\epsilon}^{k})]$
    \State $\bm\epsilon^{k+1}\!\gets\!\bm\epsilon^k-\mathcal{A}(\tilde{\mathbf{x}}^{k+1}-\tilde{\mathbf{x}}^k)+g(\tilde{\mathbf{x}}^{k},\bm{\epsilon}^{k})$
    \State $\gamma^{k+1}\gets \gamma^0\|\tilde{\mathbf{x}}^{k+1}\|_2$
    \If{convergence$(\tilde{\mathbf{x}},\bm\epsilon,\epsilon_0)$} \textbf{break} \EndIf
\EndFor
\State \textbf{return} $\tilde{\mathbf{x}}^{\star} \gets \tilde{\mathbf{x}}^{k+1}$
\end{algorithmic}}}
\label{alg3}
\end{algorithm}

For an off-grid model, the main update steps in the augmented framework are similar to those in (\ref{update_x_step1_offgrid}) and (\ref{update_grid_gaps}), as outlined below:
\begin{align}
    \tilde{\mathbf{x}}^{k+1} =& \left(\mathbf{C}^{\mathrm H}(\bm{\beta}^{k})\mathbf{C}(\bm{\beta}^{k})+\frac{\gamma^{k}(\mathrm{a}+1)^2}{\mathrm{a}\mathrm{b}^2}\bm{\Lambda}(\tilde{\mathbf{x}}^{k})\right)^{-1} \nonumber \\ &\cdot \mathbf{C}^{\mathrm H}(\bm{\beta}^{k}) \left[\mathbf{C}(\bm{\beta}^{k})\tilde{\mathbf{x}}^{k}+g^{\dagger}(\tilde{\mathbf{x}}^{k}, \bm{\epsilon}^k,\bm{\beta}^{k})\right],
    \label{x_update_offgrid_modify}
\end{align}
where  
\begin{align}
&g^{\dagger}(\tilde{\mathbf{x}}^{k},\bm{\epsilon}^{k}, \bm{\beta}^{k}) =\frac{(\mathrm{a}+1)^2\Re({\overline{\mathbf{y}}})} {\mathrm{b}\exp\left(\mathrm{b} \Re({\overline{\mathbf{y}}})\odot\Re \left(\mathbf{C}(\bm{\beta}^{k})\tilde{\mathbf{x}}^k+\bm{\epsilon}^k \right) \right)+\mathrm{a}\mathrm{b}} \nonumber \\ 
    &+ j\frac{(\mathrm{a}+1)^2\Im({\overline{\mathbf{y}}})} {\mathrm{b}\exp\left(\mathrm{b}\Im({\overline{\mathbf{y}}})\odot \Im \left(\mathbf{C}(\bm{\beta}^{k})\tilde{\mathbf{x}}^k+\bm{\epsilon}^k \right) \right)+\mathrm{a}\mathrm{b}}.
\end{align}
The update for $\bm{\epsilon}$ is given by:
\begin{equation}
    \bm{\epsilon}^{k+1} = \bm{\epsilon}^{k}-\mathbf{C}(\bm{\beta}^{k})\left (\tilde{\mathbf{x}}^{k+1} - \tilde{\mathbf{x}}^{k} \right)+g^{\dagger}(\tilde{\mathbf{x}}^{k},\bm{\epsilon}^k,\bm{\beta}^{k}),
    \label{epsi_update_offgrid_modify}
\end{equation}
and the update for grid gaps $\bm{\beta}$ follows:
\begin{align}
    \bm{\beta}^{k+1} &= \left[\Re \left((\mathcal{B}^{\mathrm{H}}\mathcal{B})^{*}(\tilde{\mathbf{x}}^{k+1}\!-\tilde{\mathbf{x}}^{k} )(\tilde{\mathbf{x}}^{k+1}-\tilde{\mathbf{x}}^{k} )^{\mathrm H}\right) \right]^{-1} \nonumber \\
    &\cdot\Re\left(\mathrm{diag}(\tilde{\mathbf{x}}^{k+1}\!-\!\tilde{\mathbf{x}}^{k})^*\mathcal{B}^{\mathrm H}\!\left[g^{\dagger}(\tilde{\mathbf{x}}^{k},\bm{\epsilon}^k,\bm{\beta}^{k})\!-\!\mathcal{A}(\tilde{\mathbf{x}}^{k+1}\!-\!\tilde{\mathbf{x}}^{k} )\right]\right).    \label{beta_update_offgrid_modify}
\end{align}
The constraint on the grid gaps, $\beta_n \in \left[-\frac{r}{2}, \frac{r}{2}\right]$, is enforced to guarantee that:
\begin{equation}
    \beta_{n}^{k+1} =
\begin{cases}
\beta_{n}^{k+1}, & \text{if } \beta_{n}^{k+1} \in \left[ -\frac{r}{2}, \frac{r}{2} \right], \\
-\frac{r}{2}, & \text{if } \beta_{n}^{k+1} < -\frac{r}{2}, \\
\frac{r}{2}, & \text{otherwise}.
\end{cases}
\label{constraint_on_gridgaps}
\end{equation}
The SBRI-X for one-bit off-grid DOA estimation is outlined in Algorithm \ref{alg4}.
\begin{algorithm}
\caption{SBRI-X for One-Bit Off-Grid DOA.}
\begin{algorithmic}[1]
\State \textbf{Input:} 
$\overline{\mathbf{y}}$, 
$\mathbf{\mathcal{A}}$,
$\mathbf{\mathcal{B}}$,
$T_{\max}$, $\alpha$, $\mathrm{a}$, $\mathrm{b}$, $\eta$, $\gamma^{0}$, error bound $\epsilon_{0}$
\State \textbf{Output:} $\tilde{\mathbf{x}}^{\star}$, $\bm{\beta}^{\star}$
\State Initialize $\tilde{\mathbf{x}}^{0} = \frac{\mathcal{A}^{\mathrm H} \overline{\mathbf{y}}}{\|\mathcal{A}^{\mathrm H} \overline{\mathbf{y}}\|_2}$, $\bm{\beta}^0 = \mathbf{0}$
\For{$k = 0$ to $T_{\max}-1$}
    \State $\tilde{\mathbf{x}}^{k+1}\!\gets\!
    \left(\mathbf{C}^{\mathrm H}(\bm{\beta}^{k})\mathbf{C}(\bm{\beta}^{k})+\frac{\gamma^{k}(\mathrm{a}+1)^2}{\mathrm{a}\mathrm{b}^2}\bm{\Lambda}(\tilde{\mathbf{x}}^{k})\right)^{-1} $
    \Statex
    $\qquad \qquad \cdot \mathbf{C}^{\mathrm H}(\bm{\beta}^{k}) \left[\mathbf{C}(\bm{\beta}^{k})\tilde{\mathbf{x}}^{k}+g^{\dagger}(\tilde{\mathbf{x}}^{k}, \bm{\epsilon}^k,\bm{\beta}^{k})\right]
    $
    \State 
    $\bm{\epsilon}^{k+1} \gets \bm{\epsilon}^{k}-\mathbf{C}(\bm{\beta}^{k})\left (\tilde{\mathbf{x}}^{k+1} - \tilde{\mathbf{x}}^{k} \right)+g^{\dagger}(\tilde{\mathbf{x}}^{k},\bm{\epsilon}^k,\bm{\beta}^{k})$
    \State $\bm\beta^{k+1}\!\gets\!\text{gridUpdate}_{\text{SBRI-X}}(\tilde{\mathbf{x}}^{k+1},\bm\epsilon^{k+1}, \tilde{\mathbf{x}}^{k}, \bm\epsilon^{k}, \mathcal{A},\mathcal{B})$, $\text{gridUpdate}_{\text{SBRI-X}}(\cdot)$ is defined in (\ref{beta_update_offgrid_modify}) and (\ref{constraint_on_gridgaps})
    \State $\gamma^{k+1}\gets \gamma^0\|\tilde{\mathbf{x}}^{k+1}\|_2$
    \If{convergence$(\tilde{\mathbf{x}},\bm\epsilon,\bm\beta,\epsilon_0)$}\ \textbf{break}\ \EndIf
\EndFor
\State \textbf{return} $\tilde{\mathbf{x}}^{\star} \gets \hat{\mathbf{x}}^{k+1}$, $\bm{\beta}^{\star} \gets \bm{\beta}^{k+1}$
\end{algorithmic}
\label{alg4}
\end{algorithm}

\vspace{-0.25em}
\section{Model-Based Neural Networks for One-Bit DOA Estimation}\label{learning-based approach}

In this section, we design model-based neural networks to address both one-bit on-grid and off-grid DOA estimation tasks. Inspired by the algorithm unrolling paradigm \cite{Algorithm_Unrolling_SPM_2021}, the iterative steps of the sparse Bayesian 
reweighted iterative algorithms are mapped to customized neural layers.

\begin{figure*}
\centering
\includegraphics[width=0.9\textwidth]{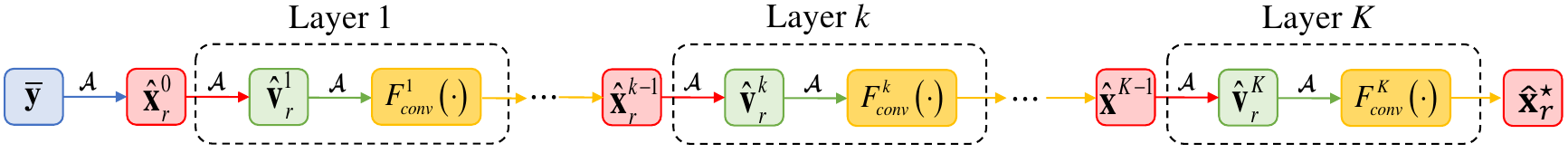}
\caption{End-to-end architecture of the SBRI-Net for the on-grid model.\label{unrolled_net_1_ongrid}}
\end{figure*}
\begin{figure*}
\centering
\includegraphics[width=0.9\textwidth]{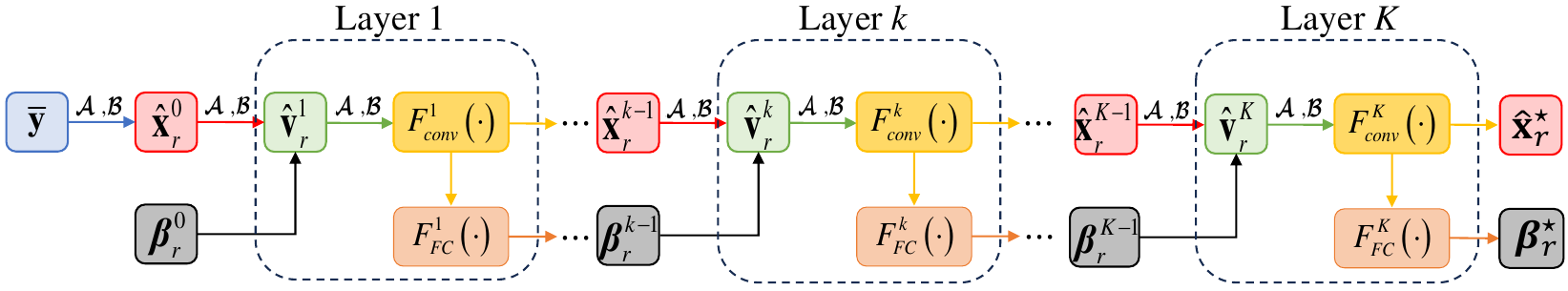}
\caption{End-to-end architecture of the SBRI-Net for the off-grid model.\label{unrolled_net_1_offgrid}}
\end{figure*}

\vspace{-0.25em}
\subsection{Inference Network Architecture }
\subsubsection{The SBRI-Net Framework} \ \\
For on-grid estimation, the network structure corresponding to Algorithm \ref{alg1} is illustrated in Figure \ref{unrolled_net_1_ongrid}. Next, we provide a detailed mathematical formulation of the network layers. Since the network parameters are defined using real values, we transform complex numbers into their real-valued counterparts. Specifically, the complex-valued measurement model (\ref{svm_model}) can be reformulated in the following real-valued form:
\begin{equation}
\overline{\mathbf{y}}_r = \mathrm{sign}\left(\mathbf{\mathcal{A}}_r\mathbf{x}_r + \mathbf{n}_r \right),
\end{equation}
where subscript $r$ is used to emphasize real-valued valuables, 
\begin{align}
\overline{\mathbf{y}}_r =& \begin{bmatrix}
\Re(\overline{\mathbf{y}})
 \\
\Im(\overline{\mathbf{y}})
\end{bmatrix} \in \mathbb{R}^{2M\times 1}, \\
\mathbf{\mathcal{A}}_r = & \begin{bmatrix}
 \Re({\mathbf{\mathcal{A}}}) & -\Im({\mathbf{\mathcal{A}}}) \\
 \Im({\mathbf{\mathcal{A}}}) & \Re({\mathbf{\mathcal{A}}}) 
\end{bmatrix} \in \mathbb{R}^{2M\times 2N}, \\
\mathbf{x}_r = &\begin{bmatrix}
\Re(\mathbf{x})
 \\
\Im(\mathbf{x})
\end{bmatrix} \in \mathbb{R}^{2N\times 1},\\
\mathbf{n}_r = &\begin{bmatrix}
\Re(\mathbf{n})
 \\
\Im(\mathbf{n})
\end{bmatrix} \in \mathbb{R}^{2M\times 1}.
\end{align}

The input to the network is the same as the initialization step in the algorithm framework but using real values $\hat{\mathbf{x}}_r^{0} = \mathcal{A}_{r}^{T} \overline{\mathbf{y}}_r$. For the $k$-th layer of the network, the input is $\hat{\mathbf{x}}_r^{k-1}$. The computation of $\hat{\mathbf{v}}_r^k$ is given by:
\begin{equation}
    \hat{\mathbf{v}}_r^k = \overline{\mathbf{y}}_r \odot \left(\mathbf{d}_{r}^{k}-\mathrm{I}'(\mathbf{d}_{r}^{k}) \right)
    \label{compute_vrk}
\end{equation}
with $    \mathbf{d}_{r}^{k} = \overline{\mathbf{y}}_r \odot (\mathbf{\mathcal{A}}_r\mathbf{x}_r^{k-1})$.
Instead of using the update formula in (\ref{step_x_update}), the CNN structures are utilized to avoid matrix inversion operations. Consequently, $\mathbf{x}$ is updated as:
\begin{equation}
\hat{\mathbf{x}}_r^{k} = F_{conv}^{k}\left(  \mathbf{\mathcal{A}}_{r}^{\mathrm T}\hat{\mathbf{v}}_r^{k-1};\mathbf{W}_{conv}^{k}, \mathbf{b}_{conv}^{k}\right),
\label{forward_cnn_one_layer}
\end{equation}
where $F_{conv}^{k}$ represents the CNN structure in the $k$-th layer, and $\mathbf{W}_{conv}^{k}$ and $\mathbf{b}_{conv}^{k}$ denote its corresponding weights and biases, respectively. Assuming that the CNN structure consists of $J$ 1D convolutional layers, the total number of convolutional layers is $KJ$, in our experiments we set $J=4$. 

Following the same design paradigm, we construct a similar network architecture for off-grid DOA estimation, as illustrated in Figure \ref{unrolled_net_1_offgrid}. The network initialization follows the same step as outlined in line \ref{init_step_alg2} of Algorithm \ref{alg2}. The update for $\hat{\mathbf{v}}^{k}$ is computed as:
\begin{equation}
    \hat{\mathbf{v}}_r^k = \overline{\mathbf{y}}_r \odot \left(\tilde{\mathbf{d}}_{r}^{k-1}-\mathrm{I}'(\tilde{\mathbf{d}}_{r}^{k-1}) \right),
    \label{update_v_sbri_offgrid}
\end{equation}
where
\begin{align}
\tilde{\mathbf{d}}_{r}^{k-1} = \overline{\mathbf{y}}_r \odot \left (\left[\mathbf{\mathcal{A}}_r+\mathbf{\mathcal{B}}_r\mathrm{diag}(\bm{\beta}_{r}^{k-1}) \right]\hat{\mathbf{x}}_r^{k-1} \right).
\end{align}
The signal coefficients $\hat{\mathbf{x}}_r^{k}$ and grid gaps $\bm{\beta}_r^{k}$ are updated as:
\begin{align}
\hat{\mathbf{x}}_r^{k}  &= F_{conv}^{k}\left(  \left[\mathbf{\mathcal{A}}_{r}+\mathbf{\mathcal{B}}_{r}\mathrm{diag}(\bm{\beta}_r^{k-1}) \right]^{\mathrm T}\hat{\mathbf{v}}_r^{k};\mathbf{W}_k, \mathbf{b}_k\right) ,\\
\bm{\beta}_r^{k} &= F_{FC}^{k}\left(\hat{\mathbf{x}}_r^{k};\mathbf{W}_{FC}^{k},\mathbf{b}_{FC}^{k}\right),
\label{fclayer_grid_offset_sbri}
\end{align}
where $\mathbf{W}_{FC}^{k}$ and $\mathbf{b}_{FC}^{k}$ denote the weights and bias, respectively, for the $k$-th fully connected layer. 
Instead of updating the grid gaps as described in line \ref{grid_gaps_update} of Algorithm \ref{alg2}, we utilize fully connected layers to produce the grid gap estimates for the next layer. 

\subsubsection{The SBRI-X-Net Framework} 
\begin{figure*}
\centering
\includegraphics[width=0.9\textwidth]{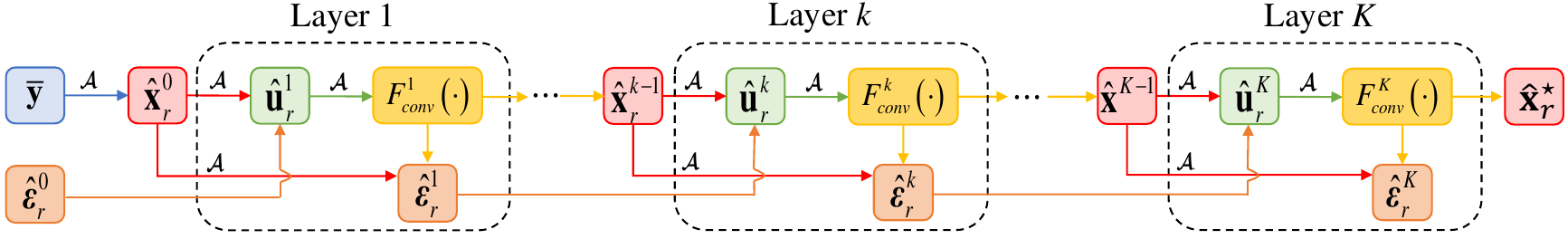}
\caption{End-to-end architecture of the SBRI-X-Net for the on-grid model.\label{unrolled_net_2_ongrid}}
\end{figure*}

\ \\
The end-to-end structure of the augmented network is shown in Figure \ref{unrolled_net_2_ongrid}. Specifically, the initialization step follows the same as that in the vanilla network. For the $k$-th layer in the network, $\hat{\mathbf{u}}_r$ is updated as:
\begin{equation}
\hat{\mathbf{u}}_r^{k} = \mathbf{\mathcal{A}}_r\hat{\mathbf{x}}_r^{k-1}+g_r(\hat{\mathbf{x}}_r^{k-1},\hat{\bm{\epsilon}}_r^{k-1}),
\end{equation}
where
\begin{equation}
g_r(\hat{\mathbf{x}}_r^{k-1},\hat{\bm{\epsilon}}_r^{k-1}) = \frac{(\mathrm{a}_k+1)^2\overline{\mathbf{y}}_r}{\mathrm{b}_k\exp\left(\mathrm{b}_k\overline{\mathbf{y}}_r\odot \left[\mathbf{\mathcal{A}}\hat{\mathbf{x}}_r^{k-1}+\hat{\bm{\epsilon}}_r^{k-1} \right] \right)+\mathrm{a}_k\mathrm{b}_k}.
\end{equation}
Noted that $\mathrm{a}_k$ and $\mathrm{b}_k$ are both learnable parameters during the end-to-end training. 
The updates for $\hat{\mathbf{x}}^{k}$ and $\hat{\bm{\epsilon}}_r^{k}$ are given by:
\begin{align}
\hat{\mathbf{x}}_r^{k} = & F_{conv}^{k}\left(  \mathbf{\mathcal{A}}_{r}^{\mathrm T}\hat{\mathbf{u}}_r^{k};\mathbf{W}_{conv}^{k}, \mathbf{b}_{conv}^{k}\right), \\
\hat{\bm{\epsilon}}_r^{k} = &\bm{\epsilon}^{k-1}-\mathbf{\mathcal{A}}_r \left [\hat{\mathbf{x}}_r^{k} - \hat{\mathbf{x}}_r^{k-1} \right ] +g_r(\hat{\mathbf{x}}_r^{k-1},\hat{\bm{\epsilon}}_r^{k-1}).
\end{align}

\begin{figure*}
\centering
\includegraphics[width=0.9\textwidth]{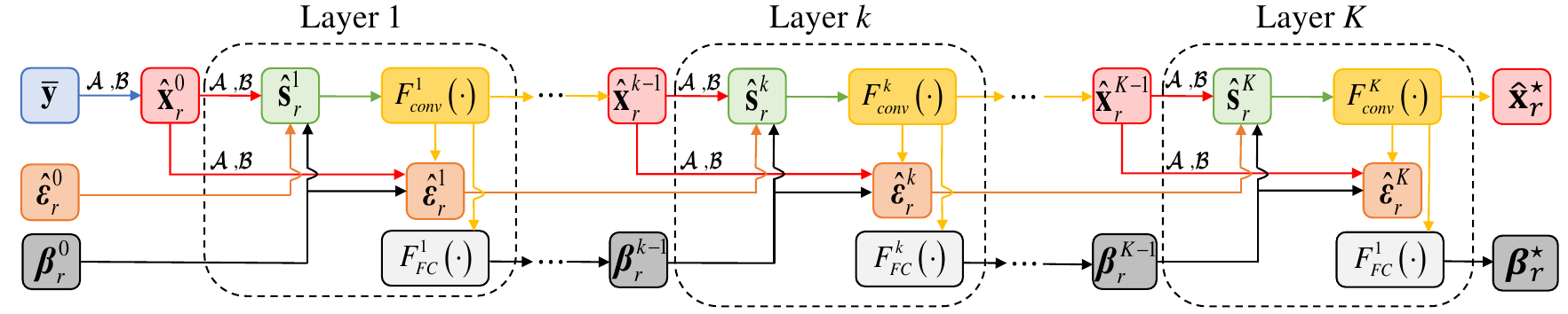}
\caption{End-to-end architecture of the SBRI-X-Net for the off-grid model.\label{unrolled_net_2_offgrid}}
\vspace{-1em}
\end{figure*}
Similarly, the end-to-end architecture of the augmented network for the off-grid model is illustrated in Figure \ref{unrolled_net_2_offgrid}. In the $k$-th layer, the computation of $\hat{\mathbf{s}}^k$ is formulated as:
\begin{align}
    \hat{\mathbf{s}}^k = \left[\mathbf{\mathcal{A}}_{r}+\mathbf{\mathcal{B}}_{r}\mathrm{diag}(\bm{\beta}_r^{k-1}) \right]\hat{\mathbf{x}}_r^{k-1}+g_r^{\dagger}(\hat{\mathbf{x}}_r^{k-1},\hat{\bm{\epsilon}}_{r}^{k-1},\bm{\beta}_r^{k-1}).
\end{align}
Finally, $\hat{\mathbf{x}}_r^{k}$, $\hat{\bm{\epsilon}}_r$ and the grid gaps $\bm{\beta}_r^{k}$ are updated as:
\begin{align}
\hat{\mathbf{x}}_r^{k} = & F_{conv}^{k}\left(  \left[\mathbf{\mathcal{A}}_{r}+\mathbf{\mathcal{B}}_{r}\mathrm{diag}(\bm{\beta}_r^{k-1}) \right]^{\mathrm T}\hat{\mathbf{s}}_r^{k-1};\mathbf{W}_k, \mathbf{b}_k\right), \\
\hat{\bm{\epsilon}}_r^{k} = & \bm{\epsilon}^{k-1}-\left[\mathbf{\mathcal{A}}_{r}+\mathbf{\mathcal{B}}_{r}\mathrm{diag}(\bm{\beta}_r^{k-1}) \right] \left[\hat{\mathbf{x}}_r^{k} - \hat{\mathbf{x}}_r^{k-1} \right] \nonumber\\
&+g_r^{\dagger}(\hat{\mathbf{x}}_r^{k-1},\hat{\bm{\epsilon}}_r^{k-1},\hat{\bm{\beta}}_{r}^{k-1}), \\
\bm{\beta}_r^{k} = & F_{FC}^{k}\left(\hat{\mathbf{x}}_r^{k-1};\mathbf{W}_{FC}^{k},\mathbf{b}_{FC}^{k}\right).
\end{align}

\vspace{-1.0em}

\subsection{Data Generation and Labeling}\label{generate_data}
We generate training datasets for on-grid and off-grid DOA estimation tasks, respectively using two SLA configurations with half-wavelength element spacing: an 18-element array at $\frac{\lambda}{2}[0, 1, 2, 3, 4, 7, 8, 9, 10, 11, 12, 13, 14, 15, 16, 17, 18, 19]$ and a 10-element array at $\frac{\lambda}{2}[0, 3, 4, 5, 6, 7, 11, 16, 18, 19]$. The number of signals is set to 2, and the array field of view (FOV) is defined as $[-60^{\circ}, 60^{\circ}]$. For on-grid training data generation, the target DOAs are randomly sampled as integer degrees within the FOV. For off-grid training data generation, the angular space is discretized into uniform grids with a fixed grid size of $2^{\circ}$. The off-grid gaps corresponding to the signals are randomly generated following a uniform distribution $U\left(-1^{\circ} , 1^{\circ}\right)$. Reflection coefficients of the two signals are generated as random complex numbers, with their real and imaginary parts uniformly distributed in $U\left(0.5, 1\right)$. Denoting the ground truth of the $n$-th DOA as set ${\mathbb G}_n$, the signals are labeled according to
\begin{equation}
\mathbf{s}_{n}^{\star}  =
\begin{cases}
|s_k|, & \text{if } \theta_k \in \mathbb G_n, \\
0, & \text{otherwise,}
\end{cases}
\end{equation}
and off-grid gaps are labeled as:
\begin{equation}
\beta_{n}^{\star} = \begin{cases}
|\beta_k|, & \text{if } \theta_k \in \mathbb G_n, \\
0, & \text{otherwise.}
\end{cases}
\end{equation}
We randomly generate $1,000,000$ samples across input SNR levels ranging between $0$ dB and $30$ dB in $5$ dB increments. $90\%$ of the dataset is used for training and the remaining $10\%$ is used for validation.

\vspace{-1em}
\subsection{Loss Functions}
The loss function used in the first stage is the binary cross-entropy (BCE) loss, defined as:
\begin{equation}
    \mathcal{L}_1(\hat{\mathbf{x}}, \mathbf{s}^{\star}) = - \frac{1}{N}\sum_{i=1}^{N} \left [ s_{i}^{\star}\cdot\mathrm{log}\hat{x}_i + (1-s_{i}^{\star})\cdot\mathrm{log}(1-\hat{x}_i) \right].
\end{equation}
In the second stage, we apply a combination of the mean squared error (MSE) loss and the BCE loss, and the total loss function is defined as:
\begin{align} \mathcal{L}_2(\hat{\mathbf{x}}, \mathbf{s}^{\star};{\bm{\beta}},\bm{\beta}^{\star}) =  \mathcal{L}_1(\hat{\mathbf{x}}, \mathbf{s}^{\star})  + \frac{1}{N}\sum_{i=1}^{N} \left ({\beta}_i - \beta_{i}^{\star} \right )^{2}.
\end{align}

\vspace{-1.5em}

\subsection{Network Training}
Our training methodology differs between the on-grid and off-grid models. For the on-grid model, we implement end-to-end training with $100$ epochs. In contrast, the off-grid model undergoes a two-stage training process with totaling $200$ epochs. In the first stage, only the signal coefficient update layers are trained using the first $100$ epochs, while the off-grid gap estimation layers remain frozen. This isolates the learning of signal coefficients, ensuring robust initial representations without off-grid interference. In the second stage, all layers are unfrozen and trained end-to-end using the left $100$ epochs, allowing the model to jointly optimize both signal coefficient updates and off-grid gap estimations. Both models use a batch size of $64$ and the Adam optimizer with a learning rate of $10^{-3}$. Training was conducted on a workstation with an Intel Core i9-9820X CPU and dual NVIDIA RTX 2080 GPUs.

\vspace{-1em}

\subsection{Complexity Analysis and Comparison of Different One-Bit DOA Estimation Algorithms}
The computational complexity of the proposed 1-bit on-grid and off-grid algorithms can be summarized as follows. For SBRI, initialization requires approximately $O(NM)$ floating-point operations (FLOPs). Each iteration is dominated by a matrix inversion of size $N$, leading to a per-iteration complexity of about $\frac{2}{3}N^3 + O(MN^2 + NM)$ FLOPs. Therefore, the overall complexity across $N_{\rm iter}$ iterations is $O(N_{\rm iter} N^3)$. Off-grid SBRI follows a similar pattern, as the additional grid-gap update step does not change the asymptotic cubic scaling.

For SBRI-X, both the on-grid and off-grid variants maintain the same initialization cost as SBRI, while each iteration incurs similar computational demands due to matrix inversion and auxiliary variable updates. Consequently, the overall complexity for SBRI-X also scales as $O(N_{\rm iter} N^3)$. Thus, all classical and augmented iterative algorithms exhibit cubic scaling with respect to the grid size $N$.

For the inference network SBRI-Net, the computational complexity for 1-bit on-grid DOA estimation is approximately $8NM$ FLOPs for initialization. Each of the $K$ unrolled layers then involves an $8NM$ auxiliary-vector update and $13{,}088N$ FLOPs for the convolution block $F_{conv}^{k}$ in (\ref{forward_cnn_one_layer}). The parameters associated with the input channels, $C_{in}$, output channels, $C_{out}$, and channel length, $K_{in}$, are given as $(C_{in}, C_{out}, K_{in}) = \{(1,21,21), (21,11,11), (11,5,5), (5,1,3)\}$. This yields a total complexity of $K(8NM + 13{,}088N)$, which is $O(KN)$ for moderate $M$. For off-grid SBRI-Net, initialization requires $16NM$ FLOPs, and each layer incurs $16NM$ for the auxiliary update, $13{,}088N$ for convolution, and $12N^2$ for three fully connected layers, giving $K(16NM + 13{,}088N + 12N^2)$ FLOPs in total. Thus, the asymptotic complexity is $O(KN)$ for on-grid, and $O(KN^2)$ for off-grid estimation.

For SBRI-X-Net, the complexity follows the same structure: on-grid estimation requires $K(12NM + 13{,}088N)$ FLOPs, scaling linearly with $N$, while off-grid estimation needs $K(20NM + 13{,}088N + 12N^2)$ FLOPs per $K$ layers, growing quadratically with $N$ due to the fully connected blocks.

Table \ref{Compuational_analysis_table} summarizes the complexity of the proposed methods and reference methods such as 1-bit LIKES, 1-bit SPICE, 1-bit SLIM\cite{shang2021weighted,6190383}.
\begin{table}[htbp]
  \centering
  \small
  \caption{Computational complexity of the compared DOA estimators
    ($M$ sensors, $N$ grid points, $N_{\mathrm{iter}}$ iterations, $K$ unrolled layers).}
  \label{Compuational_analysis_table}
  \begin{tabular}{@{}lcc@{}}
    \toprule
    \textbf{Algorithm} & \textbf{Total cost}  \\
    \midrule
    1-bit LIKES  
      & $\tfrac{5}{3}N_{\mathrm{iter}}M^{3} \;+\;4\,N_{\mathrm{iter}}M^{2}N$
     \\
    1-bit SPICE  
      & $\tfrac{5}{3}N_{\mathrm{iter}}M^{3} \;+\;2\,N_{\mathrm{iter}}M^{2}N$
      \\
    1-bit SLIM  
      & $\tfrac{5}{3}N_{\mathrm{iter}}M^{3} \;+\;4\,N_{\mathrm{iter}}M^{2}N$
    \\
    \addlinespace
    1-bit SBRI (on-grid)  
      & $\tfrac{2}{3}N_{\mathrm{iter}}N^{3} \;+\;2\,N_{\mathrm{iter}}MN^{2}$
   \\
    1-bit SBRI (off-grid)  
      & $N_{\mathrm{iter}}N^{3} \;+\;3\,N_{\mathrm{iter}}MN^{2}$
     \\
    \addlinespace
    \textbf{1-bit SBRI-Net (on-grid)}  
      & $K\,(8\,N\,M \;+\; 13\,088\,N)$
     \\
    \textbf{1-bit SBRI-Net (off-grid)}  
      & $K\,(16\,N\,M \;+\; 13\,088\,N \;+\; 12\,N^{2})$
      \\
    \bottomrule
  \end{tabular}
\end{table}

Table \ref{doa_feature_summary} presents a comparison of the key features of different algorithms for 1-bit DOA estimation.
\begin{table}[htbp]
  \centering
  \setlength{\tabcolsep}{6pt}          
  \renewcommand{\arraystretch}{1.15}   
  \caption{Key features comparison for different DOA estimation approaches.}
  \label{doa_feature_summary}
  \begin{tabular}{@{}lcccccc@{}}
    \toprule[1pt]
    \textbf{Algorithms} & \textbf{Snapshot} & \textbf{SLA} &\textbf{1-Bit } & \textbf{Off-Grid }  \\
    \midrule
    1-bit SLIM   & Single   & \ding{51}  & \ding{51}   & \ding{55}  \\
    1-bit SPICE  & Single  & \ding{51} & \ding{51}       & \ding{55}  \\
    1-bit LIKES & Single  & \ding{51} & \ding{51}       & \ding{55}  \\
    OGIR    & Single/Multiple  & \ding{51}   & \ding{51}   & \ding{51} \\
    1-bit on-grid SBRI   & Single  & \ding{51}   & \ding{51}   & \ding{55}  \\
    1-bit off-grid SBRI   & Single  & \ding{51}   & \ding{51}   & \ding{51}  \\
    1-bit on-grid SBRI-X   & Single  & \ding{51}   & \ding{51}   & \ding{55}  \\
    1-bit off-grid SBRI-X   & Single  & \ding{51}   & \ding{51}   & \ding{51}  \\
    1-bit SBRI-Net   & Single/Multiple  & \ding{51}   & \ding{51}   & \ding{51}  \\ 
    1-bit SBRI-X-Net   & Single/Multiple  & \ding{51}   & \ding{51}   & \ding{51}  \\
    \bottomrule[1pt]
  \end{tabular}
\end{table}

\section{Numerical Results}
\label{Numerical Results}
This section presents a comprehensive performance evaluation of SBRI, SBRI-X, SBRI-Net, and SBRI-X-Net. 
In particular, we consider sparse target signals across various SLA configurations under different SNR conditions.

To analyze the performance of the proposed on-grid and off-grid algorithm frameworks, we use root mean square error (RMSE) and hit rate as the evaluation metrics. The RMSE is computed as $\mathrm{RMSE}=\sqrt{\frac{1}{N_s{K}}\sum_{t=1}^{N_s}||\hat{\bm{\theta}}_t - \bm{\theta}^{\star}||_{2}^{2}}$, 
where $\hat{\bm{\theta}}_t$ represents the estimated DOA vector in the $t$-th test round, $K$ is the number of signals, and $N_s$ is the number of successful tests.
$\mathrm{Hit\ rate} = {N_s}/{N_t}$,
where $N_t$ is the total number of simulation trails. A DOA estimate is considered successful if the absolute errors of all estimated angles remain within a threshold of $2^\circ$; otherwise, it is regarded as a failure.

\subsection{On-Grid DOA Estimation}\label{ongrid DOA estimation numerical} 
\begin{figure*}
\centering
\subfigure[]{\includegraphics[height=0.255\textwidth]{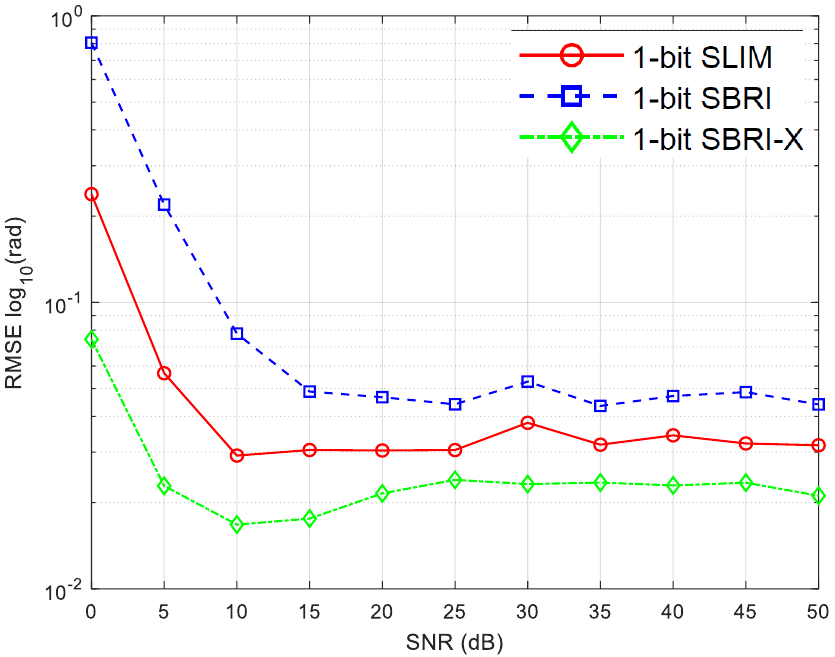}}
\subfigure[]{\includegraphics[height=0.255\textwidth]{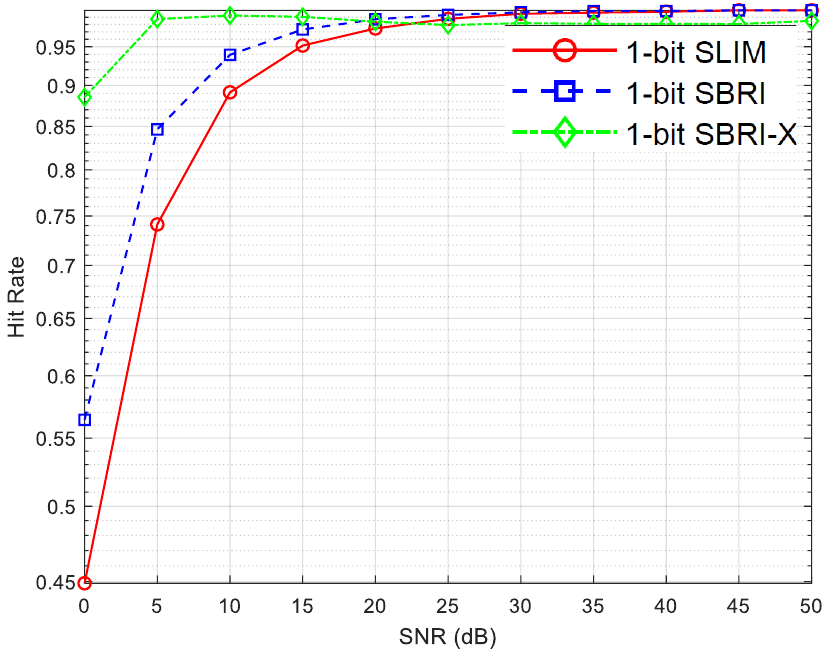}}
\subfigure[]{\includegraphics[height=0.255\textwidth]{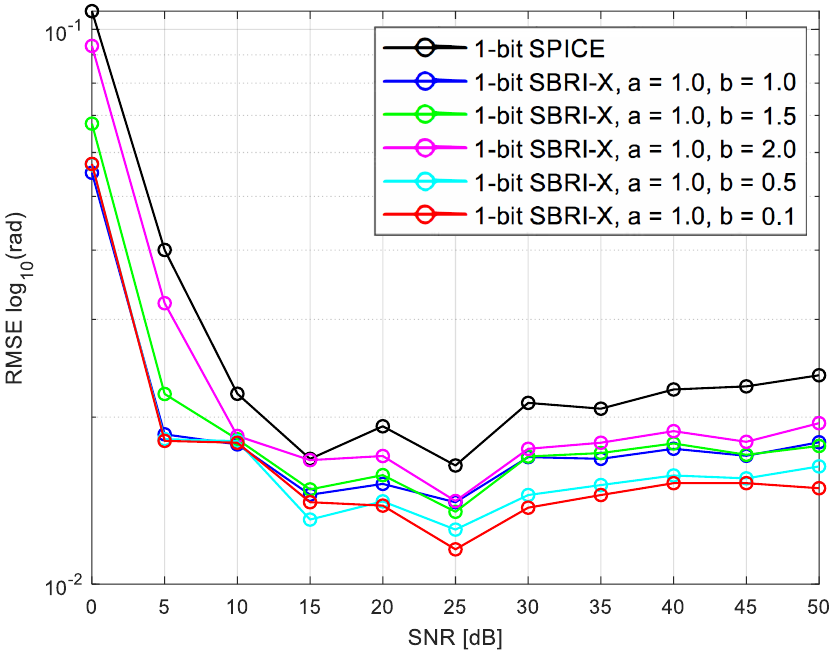}}
\subfigure[]{\includegraphics[height=0.255\textwidth]{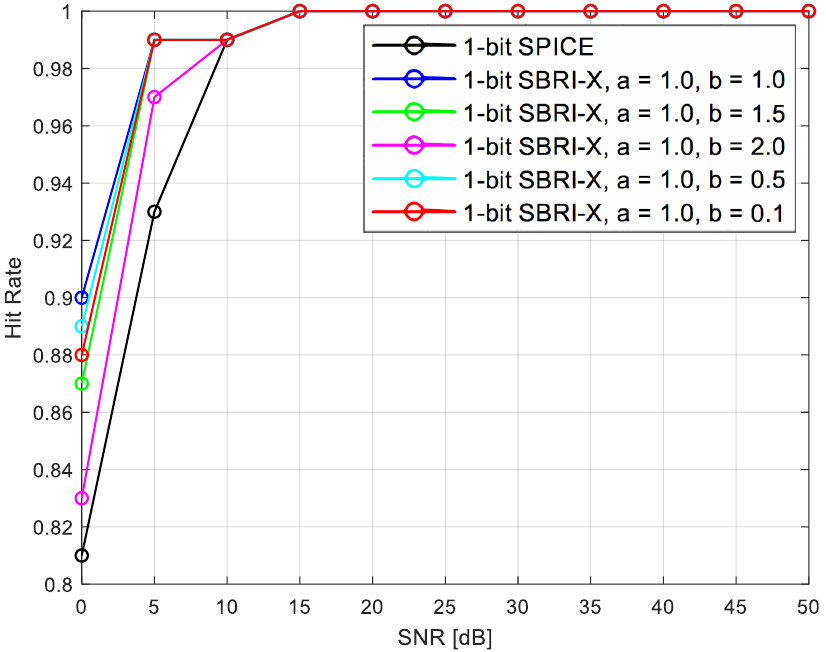}}
\subfigure[]{\includegraphics[height=0.26\textwidth]{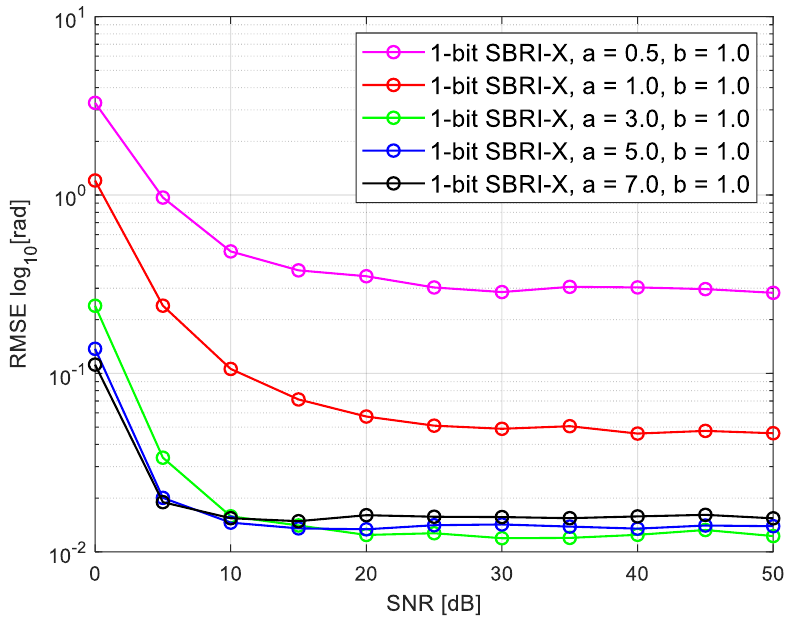}}
\subfigure[]{\includegraphics[height=0.255\textwidth]{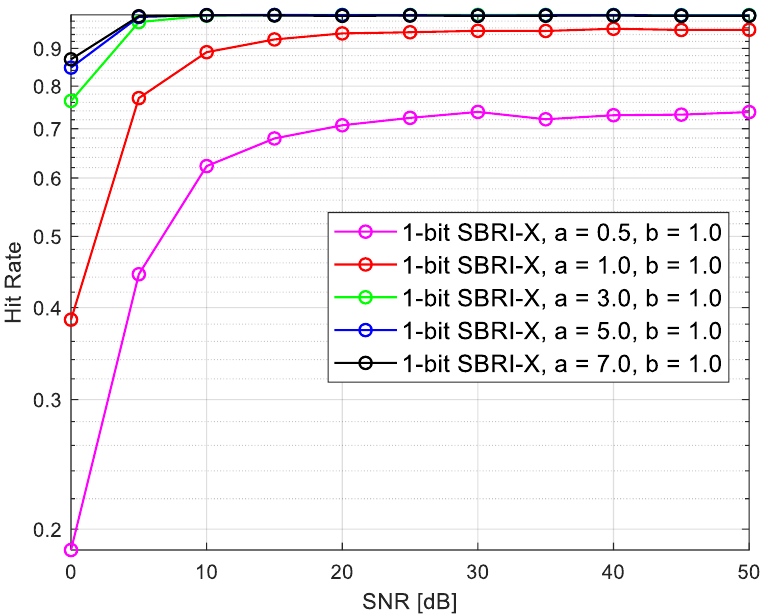}}
\vspace{-1em}
\caption{Performance comparison for on-grid DOA estimation using $18$-element SLA: (a) RMSE versus input SNR; (b) Hit rate versus input SNR; (c) RMSE versus input SNR with different selection of $\mathrm{a}$ (fixed $\mathrm{b}$); (d)  Hit rate versus input SNR with different selection of $\mathrm{a}$ (fixed $\mathrm{b}$); (e) RMSE versus input SNR with different selection of $\mathrm{b}$ (fixed $\mathrm{a}$); (f) Hit rate versus input SNR with different selection of $\mathrm{b}$ (fixed $\mathrm{a}$). }
\label{SBRI_X_RMSE1andDetect}
\vspace{-1em}
\end{figure*}

We conducted $1,000$ simulation trials using an 18-element SLA to compare the average RMSE and hit rate of the proposed algorithm against the 1-bit SLIM method \cite{shang2021weighted}. Figure \ref{SBRI_X_RMSE1andDetect} (a) presents the estimation RMSEs of the proposed 1-bit SBRI and 1-bit SBRI-X, alongside 1-bit SLIM, across different input SNR levels. Figure \ref{SBRI_X_RMSE1andDetect} (b) illustrates the estimation hit rate for these methods. The test scenario consists of two fixed targets located at $\left[-30^{\circ}, 30^{\circ}\right]$. The proposed 1-bit SBRI-X achieves the lowest RMSE across all SNR levels, outperforming both 1-bit SLIM and 1-bit SBRI. Additionally, it demonstrates the highest estimation accuracy and highest hit rate in low input SNR conditions between 0 dB and 15 dB. These results highlight the performance gains of 1-bit SBRI-X over 1-bit SBRI and validate that the chosen sigmoid likelihood function contributes significantly to performance gains, particularly in low SNR conditions.

We further investigate the impact of parameter selection on the performance of the proposed 1-bit SBRI-X algorithm. As shown in Figure \ref{example_different_likelihood}, parameter $\mathrm{b}$ controls the slopes of the likelihood functions. Figure ~\ref{SBRI_X_RMSE1andDetect} (c-d) presents the estimation curves of 1-bit SBRI-X under different $\mathrm{b}$ settings. It can be observed that reducing $\mathrm{b}$ results in lower RMSEs across various SNR levels and a higher hit rate, indicating that smaller slopes of the likelihood function enhance the  performance. Figure \ref{SBRI_X_RMSE1andDetect} (c) shows that the RMSE decreases at first and then rises as the input SNR increases. This trend can be attributed to the influence of noise and quantization error at different SNR levels. In low SNR conditions, noise is the dominant factor affecting estimation accuracy, whereas in high SNR conditions, the impact of noise diminishes, and quantization error becomes the dominant factor. This performance saturation phenomenon has also been reported in the literature, e.g., \cite{10848316}. To conduct more ablation studies, we compared estimation performance under fixed $b=1.0$ but various $\mathrm{a}$. Figure \ref{SBRI_X_RMSE1andDetect} (e) shows that increasing $\mathrm{a}$ reduces RMSE at every SNR, but the gain saturates when $\mathrm{a} \ge 3$; beyond this point the curve flattens, demonstrating a ``diminishing-returns'' effect frequently reported in hyper-parameter studies. In low-SNR regimes ($0$--$10$ dB), larger $\mathrm{a}$ gives a marked RMSE improvement, whereas at high SNR an overly larger $\mathrm{a}$ slightly degrades accuracy. Hit-rate curves in Figure \ref{SBRI_X_RMSE1andDetect} (f) mirror the same trend: increasing $a$ maximises detections. Taken together with the sensitivity to $b$ shown in Figure \ref{SBRI_X_RMSE1andDetect} (e-f), these results confirm that SBRI-X is responsive to $\left(\mathrm{a}, \mathrm{b}\right)$. Together, these ablations confirm that both $\mathrm{a}$ and $\mathrm{b}$ shape SBRI-X performance. 

\begin{figure*}
\centering
\subfigure[]{\includegraphics[width=0.32\textwidth]{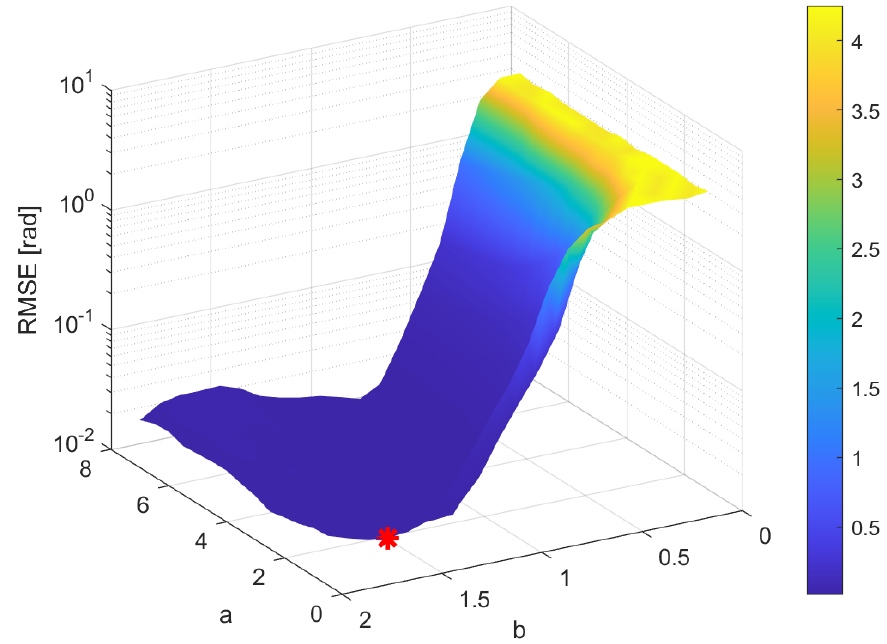}}
\subfigure[]{\includegraphics[width=0.32\textwidth]{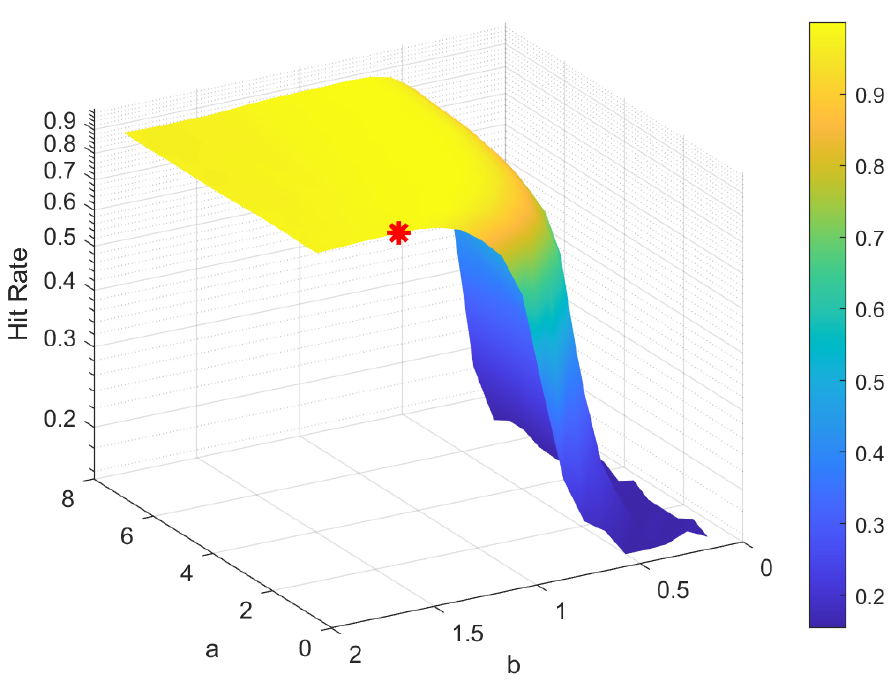}}
\subfigure[]{\includegraphics[width=0.32\textwidth]{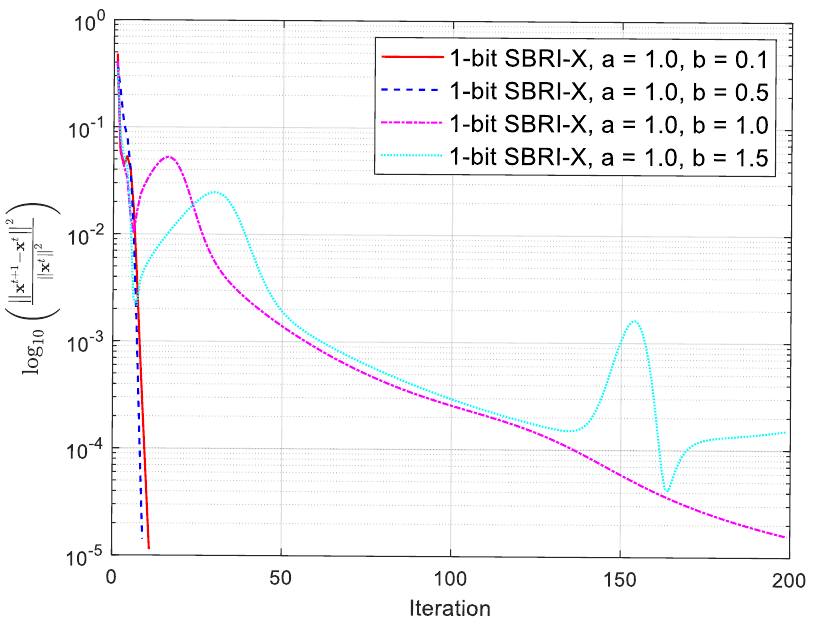}}
\vspace{-1em}
\caption{Effects of 2D tuning of the sigmoid parameters $\mathrm{a}$ and $\mathrm{b}$ on RMSE and Hit rate for on-grid DOA estimation using the proposed SBRI-X (18-element SLA) under 10 dB input SNR: (a) RMSE versus 2D grid search with different $\mathrm{a}$ and $\mathrm{b}$; (b) Hit rate versus 2D grid search with different $\mathrm{a}$ and $\mathrm{b}$; (c) Convergence curves of 1-bit SBRI-X. 
\label{rmse_hitrate_2d_effect}}
\vspace{-0.5em}
\end{figure*}

We can further optimize the parameters $\mathrm{a}$ and $\mathrm{b}$ via two-dimensional (2D) grid search. As illustrated in Figure \ref{rmse_hitrate_2d_effect}, the lowest RMSE is attained when the parameter pair $\left(\mathrm{a}, \mathrm{b}\right)$ is close to $\left(0.5, 1.7\right)$.

Additionally, Figure \ref{rmse_hitrate_2d_effect} (c) presents the convergence curves for different choices of $\mathrm{b}$. It can be observed that smaller values of $\mathrm{b}$ (e.g., $\mathrm{b} = 0.1$ or $\mathrm{b} = 0.5$) result in a faster convergence rate, reaching smaller approximation error within approximately $10$ iterations.

\begin{figure*} 
\centering
\subfigure[]{\includegraphics[height=0.25\textwidth]{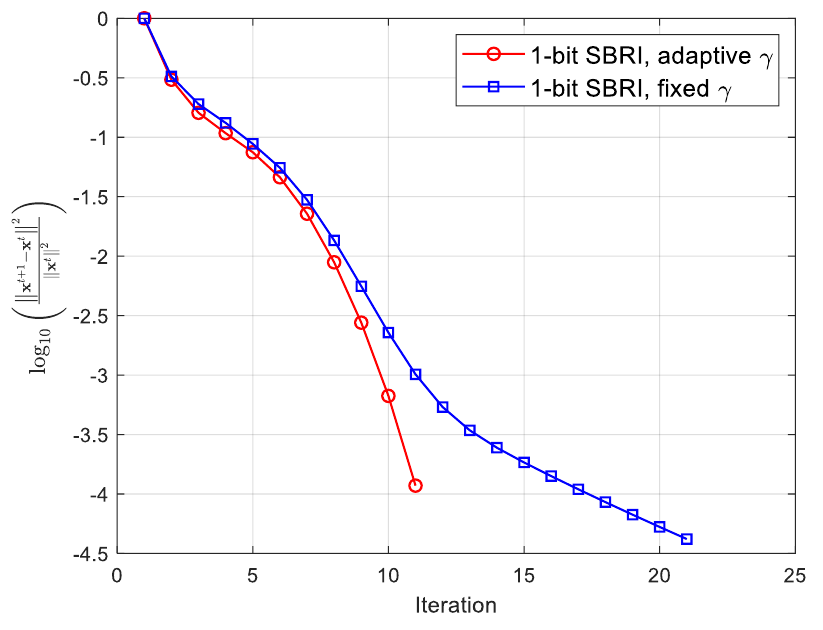}}
\subfigure[]{\includegraphics[height=0.25\textwidth]{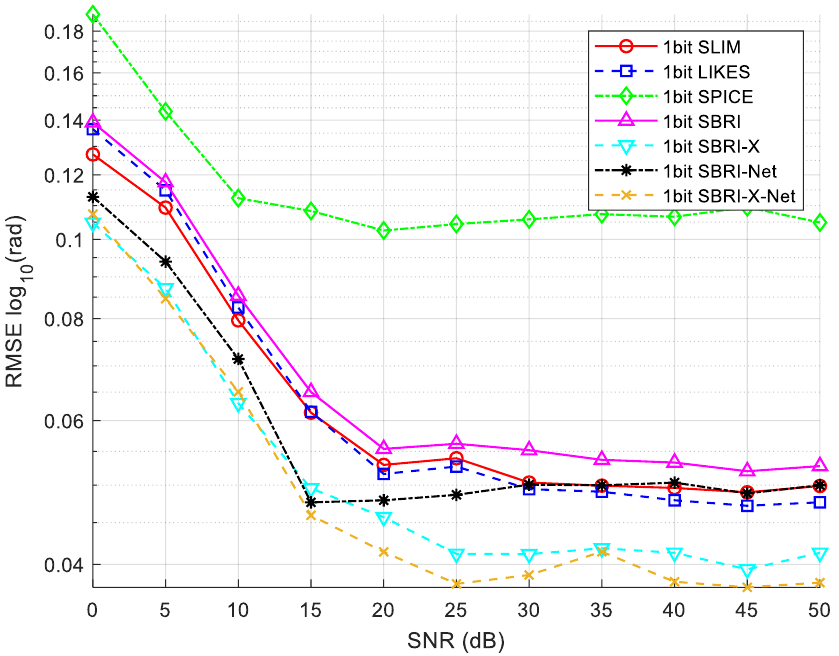}}
\subfigure[]{\includegraphics[height=0.25\textwidth]{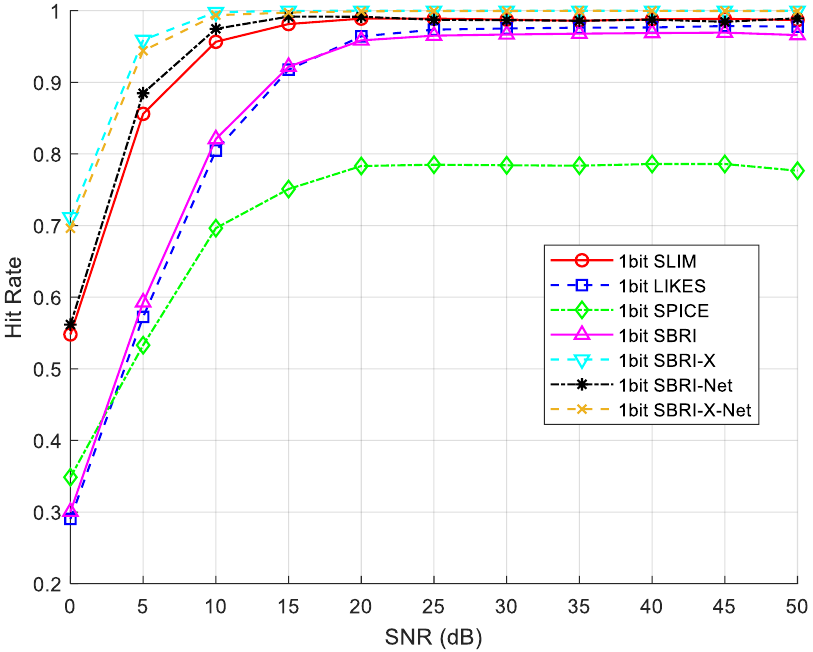}}
\vspace{-1em}
\caption{The performance of the proposed augmented algorithms and networks for on-grid DOA estimation with an 18-element SLA: (a) Convergence curves of 1-bit SBRI with different regularization settings; (b) RMSE versus input SNR; (c) Hit rate versus input SNR. 
\label{convergence_curves_sbri_reg}}
\vspace{-1em}
\end{figure*}

Note also that the choice of the regularization parameter $\gamma$ will affect the algorithm convergence speed. We compared convergence under different $\gamma$ settings and found that, as Figure~\ref{convergence_curves_sbri_reg} (a) shows, the adaptive rule in Equation~\eqref{lambda_update} accelerates convergence compared with a fixed $\gamma$.

To further evaluate the performance of the proposed SBRI-Net and SBRI-X-Net (illustrated in Figure~\ref{unrolled_net_1_ongrid} and Figure~\ref{unrolled_net_1_offgrid}), we conducted $5,000$ random trials using an 18-element SLA. These trials focused on a scenario with two fixed on-grid targets located at $\left[-10^{\circ}, 20^{\circ}\right]$ under varying SNR levels. The results were compared against optimization-based estimators, including 1-bit SPICE, 1-bit LIKES, and 1-bit SLIM \cite{shang2021weighted}. As shown in Figure \ref{convergence_curves_sbri_reg} (b) and (c),  
SBRI-X-Net achieves comparable RMSE to SBRI-X at SNRs from $0$ dB to $15$ dB, and slightly lower RMSE at $20$ dB to $50$ dB. It also maintains a comparable hit rate across various SNR conditions, including those ranging from $35$ dB to $50$ dB, which were not part of the training scenarios. This demonstrates better overall performance and a certain level of generalization ability in high SNR conditions. A similar trend is observed for SBRI-Net, further demonstrating the effectiveness of the proposed model-based networks.

\begin{figure} 
\centering
\includegraphics[width=0.4\textwidth]{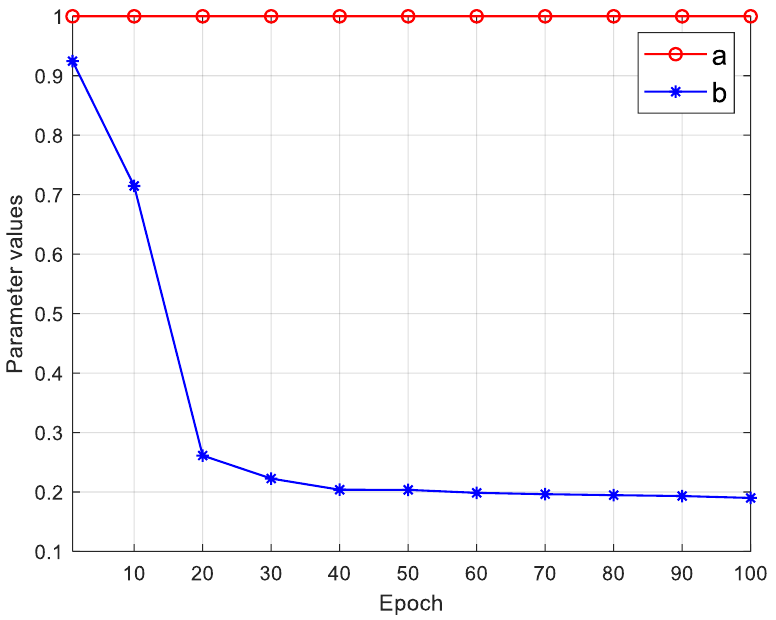}
\caption{Evolution of the learned sigmoid parameters $(\mathrm{a},\mathrm{b})$ across training epochs for SBRI-X-Net.
}
\label{a_b_epoch}
\vspace{-1em}
\end{figure}

Unlike the original SBRI-X algorithm, which requires manually chosen $a$ and $b$, SBRI-X-Net learns these parameters during training. We initialize both $a$ and $b$ to 1.0. Because SBRI-X is not strictly an unrolled network, letting both parameters float simultaneously led to numerical instabilities (NaN propagation). To avoid this, we fix $a$ and allow only $b$ to be trainable. Figure \ref{a_b_epoch} plots $b$ (and fixed $a$) over training epochs. We see that $b$ steadily decreases, mirroring the behavior observed in our manual grid searches (Figure~\ref{SBRI_X_RMSE1andDetect} (c-d)), where smaller $b$ consistently yielded better estimation performance.

\vspace{-4mm}
\subsection{Off-Grid DOA Estimation}\label{offgrid DOA estimation numerical} 
\begin{figure*}
\centering
\subfigure[]{\includegraphics[height=0.26\textwidth]{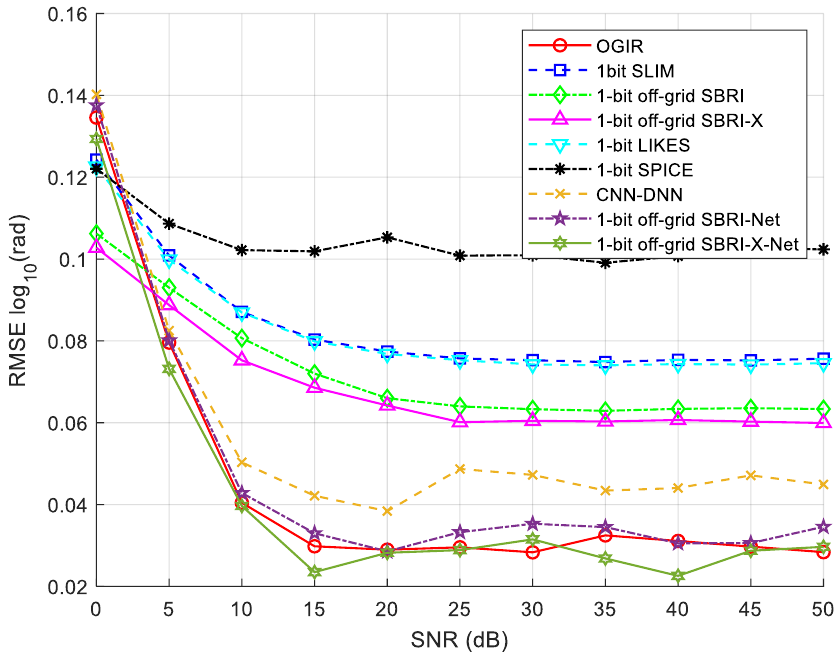}}
\subfigure[]{\includegraphics[height=0.26\textwidth]{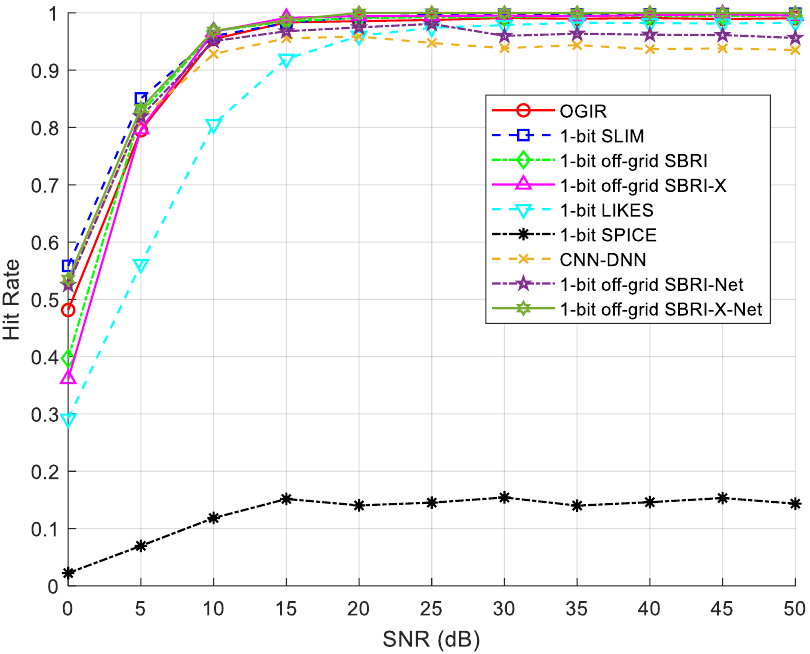}}
\subfigure[]{\includegraphics[height=0.26\textwidth]{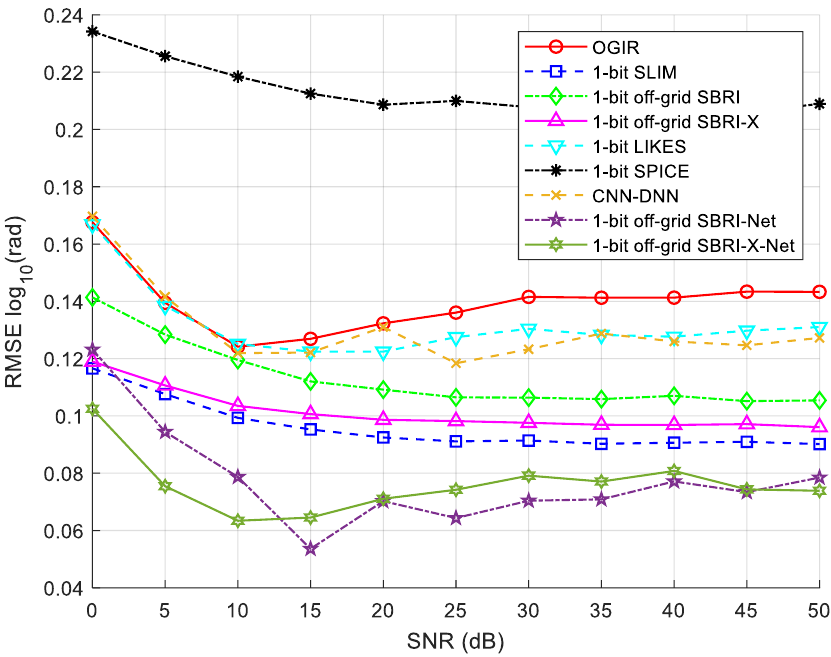}}
\subfigure[]{\includegraphics[height=0.26\textwidth]{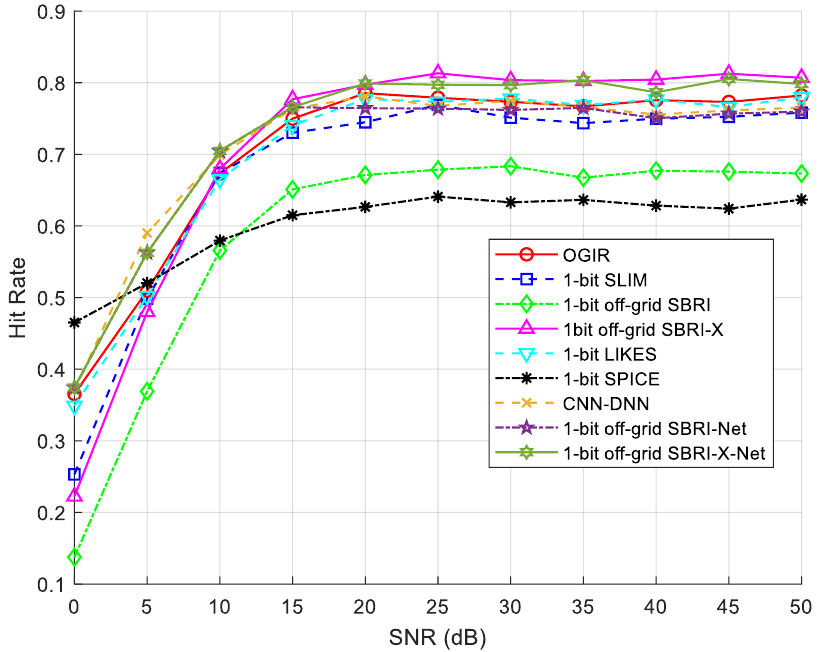}}
\subfigure[]{\includegraphics[height=0.26\textwidth]{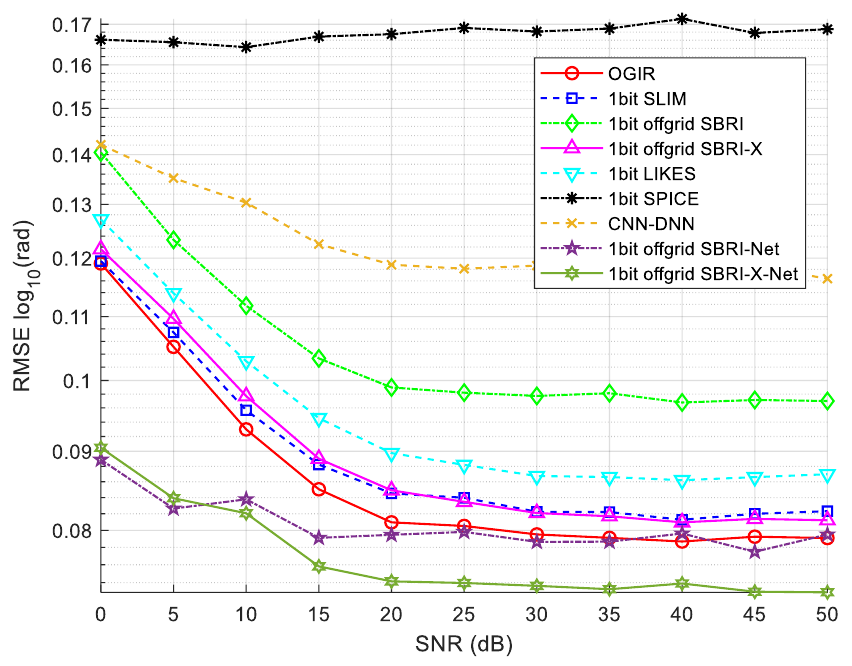}}
\subfigure[]{\includegraphics[height=0.26\textwidth]{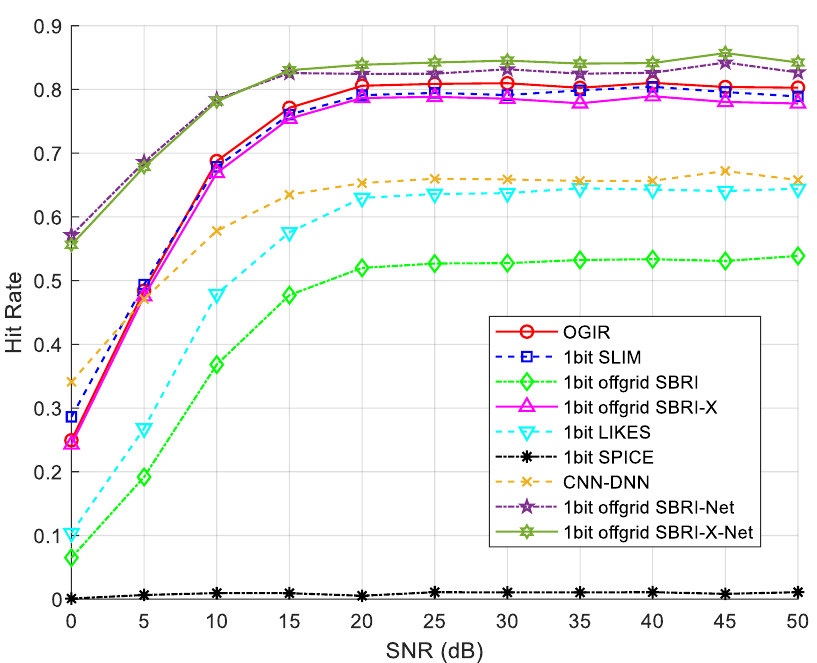}}
\vspace{-1em}
\caption{Performance comparison of SBRI-Net for off-grid DOA estimation: (a) RMSE versus input SNR for  $18$-element SLA with two targets; (b) Hit rate versus input SNR for  $18$-element SLA with two targets; (c) RMSE versus input SNR for  an $10$-element SLA with two targets; (d) Hit rate versus input SNR for  $10$-element SLA with two targets; (e) RMSE versus input SNR for $10$-element SLA with three targets; (f) Hit rate versus input SNR for $10$-element SLA with three targets.}
\label{RMSE_hitrate_offgrid_network}
\vspace{-1em}
\end{figure*}

We first conducted $5,000$ simulation trials on two fixed off-grid targets at $\left[-10.28^{\circ}, 20.56^{\circ}\right]$, using the 18-element SLA described in Section \ref{generate_data}. The proposed algorithms and networks were evaluated against several benchmarks, including the pure data-driven CNN-DNN-based method proposed in \cite{chung2021off}, and other algorithms such as OGIR \cite{10024794}, 
1-bit SLIM, 1-bit LIKES and 1-bit SPICE \cite{shang2021weighted}.
As shown in Figure~\ref{RMSE_hitrate_offgrid_network}  (a), the off-grid methods consistently outperform on-grid algorithms. Among off-grid approaches, SBRI-X-Net achieved the lowest RMSE, while SBRI-Net performed on par with OGIR. Furthermore, the proposed model-based SBRI-X-Net and SBRI-Net exhibit superior performance over the pure data-driven CNN-DNN approach. From Figure \ref{RMSE_hitrate_offgrid_network} (b), the proposed model-based approach achieves the highest hit rate across various SNR conditions, confirming the superior estimation performance of the proposed methods.

In Figure \ref{RMSE_hitrate_offgrid_network} (c-d), we further compare the performance of the proposed networks using the 10-element SLA described in Section \ref{generate_data}. It can be observed that all methods experience performance degradation on the sparser 10-element SLA compared to their results with the 18-element SLA. However, the proposed model-based algorithms still outperform the other comparison algorithms, demonstrating their robustness for SLA configurations.

To further assess robustness at the scenario containing more sources, we carried out an
additional $5{,}000$ Monte-Carlo trials with \emph{three} off--grid
targets positioned at $(-10.28^{\circ},\,0.25^{\circ},\,20.56^{\circ})$.
As summarized in Figure~\ref{RMSE_hitrate_offgrid_network} (e-f),
the trends remain consistent with the two--source experiment:
SBRI-X-Net delivers the lowest RMSE across the entire SNR
range, while SBRI-Net maintains clear gains over OGIR (especially under low input SNR) and
the pure CNN-DNN baseline.  Both networks also achieve the highest
hit rate, demonstrating that the proposed framework
scales gracefully to scenarios with more active sources.

\vspace{-0.5em}
\section{Conclusions}
\label{conclusion}

In this paper, we addressed the challenge problem of DOA estimation of off-grid signals using only a single snapshot of one-bit quantized data via a learning-based framework that synergizes model-driven sparse recovery with data-driven optimization. By reformulating DOA estimation under a MAP approach with a Laplacian-type sparsity prior, we unify on-grid and off-grid recovery while integrating a computationally efficient first-order grid approximation for practical one-bit systems with sparse arrays. The proposed SBRI and SBRI-X algorithms outperform existing model-based one-bit DOA methods in estimation accuracy across various SNR conditions and SLA configurations. Further, their neural network counterparts, SBRI-Net and SBRI-X-Net, achieve  faster convergence by unrolling iterative optimization into interpretable deep neural layers, balancing computational efficiency with theoretical guarantees. 
The simulation results validate the superiority over existing methods in resolution threshold and hardware efficiency.
This work advances scalable, high-resolution DOA estimation for next-generation sensing applications exploiting sparse arrays.

\vspace{-1em}
\begin{appendices}

\section{Proof of Theorem 1}
\label{Appendix_A}

\begin{proof}
For $\eta > 0$, the coercivity of the regularizer ensures that the sequence $\{\hat{\mathbf{x}}^k\}_{k \geq 0}$ remains bounded. Hence, according to the Bolzano-Weierstrass theorem\cite{bartle2000introduction}, there exists a convergent subsequence $\{\hat{\mathbf{x}}^{k_\jmath}\}$ such that
\begin{align}
\hat{\mathbf{x}}^{k_\jmath} \to \hat{\mathbf{x}}^{\dagger}, \quad \text{as } \jmath \to \infty.
\end{align}
By Lemma 2 (see Appendix \ref{Appendix_Lemmas}), we have
\begin{align}
\|\hat{\mathbf{x}}^{k+1} - \hat{\mathbf{x}}^k\|_2 \to 0,
\end{align}
which implies that the shifted subsequence $\{\hat{\mathbf{x}}^{k_{\jmath+1}}\}$ also converges to the limit $\hat{\mathbf{x}}^{\dagger}$. 
Let $\{\mathbf{v}^{k_{\jmath+1}}\}$ denote the corresponding subsequence of the $\mathbf{v}$-iterates. Since the sequence $\{(\hat{\mathbf{x}}^{k},\mathbf{v}^{k})\}$ is bounded and from Lemma~3 (see Supplementary Materials), the Lyapunov function is monotonically decreasing, we define the limit point:
\begin{align}
\mathbf{v}^{\dagger} := \lim_{\jmath \to \infty} \mathbf{v}^{k_{\jmath}}.
\end{align}
Letting $\jmath \to \infty$ in the first-order optimality condition associated with the $\hat{\mathbf{x}}$-update (Equation~\eqref{step_x_update}), we obtain:
\begin{align}
  \mathbf{z} + \mathcal{A}^{\mathrm H}(\mathcal{A}\hat{\mathbf{x}}^{\dagger} - \mathbf{v}^{\dagger}) = \mathbf{0},  
\end{align}
where ${\mathbf z} = [z_1,\cdots,z_N]^T$ is the gradient of the smoothed regularizer:
\begin{align}
    z_i = \frac{\gamma \hat{x}_i^{\dagger}}{(\eta + |\hat{x}_i^{\dagger}|^2)^{1 - \frac{\alpha}{2}}}, \ i=1,...,N,
\end{align}
and $\hat{\mathbf{x}}^{\dagger}$ is the limit point of the surrogate objective in the $\hat{\mathbf{x}}$-update step.
Furthermore, Lemma~2 in the Appendix \ref{Appendix_Lemmas} implies that the function $\mathcal{L}(\hat{\mathbf{x}}^k, \gamma)$ is monotonically decreasing and bounded below, so it converges. In particular, we have 
\begin{align}
\mathcal{L}(\hat{\mathbf{x}}^{\dagger}, \gamma) \leq \mathcal{L}(\hat{\mathbf{x}}^0, \gamma) \leq \|\hat{\mathbf{x}}^0\|_\alpha^2 + N,
\end{align}
where $\|\hat{\mathbf{x}}\|_\alpha^2 := \sum_{i=1}^{N}(|x_i|^2 + \eta)^{\alpha/2}$. This leads to the following bound on the residual:
\begin{align}
\|\mathcal{A}\hat{\mathbf{x}}^{\dagger} - \mathbf{v}^{\dagger}\|_2 \leq \sqrt{2\,\mathcal{L}(\hat{\mathbf{x}}^{\dagger}, \gamma)} \leq \sqrt{2(\|\hat{\mathbf{x}}^0\|_\alpha^2 + N)}.
\end{align}
Now consider the recovery error with respect to a ground-truth $s$-sparse signal $\hat{\mathbf{x}}^\diamondsuit$ with support set $S := \mathrm{supp}(\hat{\mathbf{x}}^\diamondsuit)$. Let $S^\dagger$ be the index set corresponding to the $s$ largest entries (in magnitude) of $\hat{\mathbf{x}}^{\dagger}$. Using standard sparse recovery arguments and the RIP assumption $\delta_{2s} < 1$, we have:
\begin{align}
&\|\hat{\mathbf{x}}^{\dagger} - \hat{\mathbf{x}}^\diamondsuit\|_2
\le \|(\hat{\mathbf{x}}^{\dagger} - \hat{\mathbf{x}}^\diamondsuit)_{S \cup S^\dagger}\|_2 + \|(\hat{\mathbf{x}}^{\dagger})_{(S \cup S^\dagger)^c}\|_2 \nonumber \\
&\le \frac{1}{\sqrt{1 - \delta_{2s}}} \|\mathcal{A}(\hat{\mathbf{x}}^{\dagger} - \hat{\mathbf{x}}^\diamondsuit)\|_2 + \frac{1}{\sqrt{1 - \delta_{2s}}} \|(\hat{\mathbf{x}}^{\dagger})_{(S \cup S^\dagger)^c}\|_2 \nonumber \\
&\le \frac{1}{\sqrt{1 - \delta_{2s}}} \|\mathcal{A} \hat{\mathbf{x}}^{\dagger} - \mathbf{v}^{\dagger} \|_2 + \left( \frac{\|\mathcal{A}\|_2}{\sqrt{1 - \delta_{2s}}} + 1 \right) \sigma_s(\hat{\mathbf{x}}^{\dagger})_2,
\end{align}
where $\sigma_s(\hat{\mathbf{x}}^{\dagger})_2 := \inf_{\|\mathbf{y}\|_0 \le s} \|\hat{\mathbf{x}}^{\dagger} - \mathbf{y}\|_2$ denotes the best $s$-sparse approximation error.
Substituting the earlier bound on the residual norm yields:
\begin{align}
\|\hat{\mathbf{x}}^{\dagger} - \hat{\mathbf{x}}^\diamondsuit\|_2 \le C_1 + C_2 \sigma_s(\hat{\mathbf{x}}^{\dagger})_2,
\end{align}
where $C_1 := \frac{1}{\sqrt{1 - \delta_{2s}}} \sqrt{2(\|\hat{\mathbf{x}}^0\|_\alpha^2 + N)}$ and $C_2 := \frac{\|\mathcal{A}\|_2}{\sqrt{1 - \delta_{2s}}} + 1$.
This completes the proof.
\end{proof}

\section{Proof of Theorem 2}
\label{Appendix_B}

\begin{proof}
\label{Theorem_2}
For $\eta>0$ and $\gamma>0$, the boundedness of the sequence $\{\tilde{\mathbf{x}}^{k}\}_{k\ge0}$ guarantees a subsequence
$\{\tilde{\mathbf{x}}^{k_{\jmath}}\}$ converging to some limit point
$\tilde{\mathbf{x}}^{\dagger}$.  
Since Equation (116) of Lemma~4 in the Appendix \ref{Appendix_Lemmas}   asserts $\|\tilde{\mathbf{x}}^{k+1}-\tilde{\mathbf{x}}^{k}\|_2\!\to 0$,
the shifted subsequence $\{\tilde{\mathbf{x}}^{k_{\jmath+1}}\}$ also converges to the same
$\tilde{\mathbf{x}}^{\dagger}$.

Substituting $\tilde{\mathbf{x}}^{k_{\jmath}}$ and $\tilde{\mathbf{x}}^{k_{\jmath+1}}$ for
$\tilde{\mathbf{x}}^{k}$ and $\tilde{\mathbf{x}}^{k+1}$, respectively, in the
first-order derivative of the objective function in
Equation (43) in the manuscript and letting $\jmath \!\to\!\infty$ yields
\begin{align}
\mathbf{z}
\;-\;\,\mathbf{\mathcal{A}}^{\rm H}\!\bigl(\mathbf{g}(\tilde{\mathbf{x}}^{\dagger},\tilde{\mathbf{\epsilon}}^{\dagger})\bigr)
= \mathbf{0},
\end{align}
where ${\mathbf z} = [z_1,\cdots,z_N]^{\mathrm T}$, $z_i=
\frac{\gamma\,\tilde{x}^{\dagger}_i}      {\bigl(\eta+|\tilde{x}^{\dagger}_i|^{2}\bigr)^{1-\tfrac{\alpha}{2}}}, \ 1 \le i \le N,$ and  $\tilde{\mathbf{x}}^{\dagger}$ is the limit point of the objective in Equation (43).

Applying Lemma~4 in the Appendix \ref{Appendix_Lemmas} gives
\begin{align}
\bar{\mathcal{L}}_{\rm ongrid}^{\diamond}\left(\tilde{\mathbf{x}}^{\dagger}, \bm{\epsilon}^{\dagger},\gamma\right)\!
\;&\le\;
\bar{\mathcal{L}}_{\rm ongrid}^{\diamond}\left(\tilde{\mathbf{x}}^{k_i}, \bm{\epsilon}^{\dagger},\gamma\right)
\; \nonumber\\
&\le\;
\bar{\mathcal{L}}_{\rm ongrid}^{\diamond}\left(\tilde{\mathbf{x}}^{0}, \bm{\epsilon}^{\dagger},\gamma\right) \nonumber\\
&\le \;\|\tilde{\mathbf{x}}^{0}\|_{\alpha}^{2}+N.
\end{align}
Consequently, it implies that
\begin{align}
\bigl\lVert \mathbf{g}\bigl(\tilde{\mathbf{x}}^{\dagger},
\tilde{\bm{\epsilon}}^{\dagger}\bigr)\bigr\rVert_{2} &\le\;
\sqrt{\frac{2(a+1)}{a b^{2}}\,
      \bar{\mathcal{L}}_{\rm ongrid}^{\diamond}\!\bigl(
      \tilde{\mathbf{x}}^{\dagger},\bm{\epsilon}^{\dagger},\gamma\bigr)} \nonumber\\
&\le
\sqrt{\frac{2(\mathrm{a}+1)}{\mathrm{a} \mathrm{b}^{2}}\,
\bigl(\lVert\tilde{\mathbf{x}}^{\diamondsuit }\rVert_{\alpha}^{2}+N\bigr)} .
\end{align}
Let $S$ be the index set of non-zero entries of $\tilde{\mathbf{x}}^\diamondsuit$ and
$S^\dagger$ be the set of the $s$ largest entries (in magnitude) of
$\tilde{\mathbf{x}}^{\dagger}$.  
Since $\|\tilde{x}^\diamondsuit \|_0\le s$ ($s$-sparse), we have
\begin{align}
\bigl\|\tilde{\mathbf{x}}^{\dagger}-\tilde{\mathbf{x}}^\diamondsuit \bigr\|_2 
 \le&
   \bigl\|(\tilde{\mathbf{x}}^{\dagger}-\tilde{\mathbf{x}}^\diamondsuit )_{S\cup S^\dagger}\bigr\|_2
   +\bigl\|(\tilde{\mathbf{x}}^{\dagger})_{(S\cup S^\dagger)^{c}}\bigr\|_2 \nonumber\\
 \le&
   \frac{1}{\sqrt{1-\delta_{2s}}}\,
   \sqrt{\frac{2(\mathrm{a}+1)}{\mathrm{a} \mathrm{b}^{2}}\!\bigl(\|\tilde{\mathbf{x}}^{\diamondsuit }\|_{\alpha}^{2}+N\bigr)} \nonumber\\
   &+\Bigl(\frac{1}{\sqrt{1-\delta_{2s}}}\,\|\mathbf{\mathcal{A}}\|_2+1\Bigr)
   \sigma_s\!\bigl(\tilde{\mathbf{x}}^{\dagger}\bigr)_2 \nonumber\\ 
= & C_{1}' + C_{2}' \sigma_s\!\bigl(\hat{\mathbf{x}}^{\dagger}\bigr)_2.
\end{align}
This completes the proof of Theorem 2.
\end{proof}

\section{Lemmas and Proofs}
\label{Appendix_Lemmas}
\begin{lemma} \label{lem1}
Let $0<\alpha\le 1$, $\eta\ge 0$, $\gamma\ge 0$. For any $x,y\in\mathbb{C}$,
\begin{align}
&\frac{\gamma}{\alpha}( \eta + |x|^{2})^{\!\alpha/2}
-
\frac{\gamma}{\alpha}(\eta + |y|^{2})^{\!\alpha/2}
-
\frac{\gamma\,\Re\!\big\{(x-y)\,\overline{y}\big\}}{(\eta + |x|^{2})^{1-\frac{\alpha}{2}}}
\ge \nonumber \\
&\frac{\gamma\,|x-y|^{2}}{2(\eta + |x|^{2})^{1-\frac{\alpha}{2}}}.
\label{inequality_1}
\end{align}
Here, $\overline{y}$ denotes the complex conjugate of $y$.
\end{lemma}

\begin{proof}
Let $x=x_{\mathrm{R}}+j x_{\mathrm{I}}$ and $y=y_{\mathrm{R}}+j y_{\mathrm{I}}$, and define
$u=[x_{\mathrm{R}},\,x_{\mathrm{I}}]^{\mathrm T}$, $v=[y_{\mathrm{R}},\,y_{\mathrm{I}}]^{ \mathrm T}\in\mathbb{R}^{2}$.
Then $|x|^{2}=\|u\|^{2}$, $|y|^{2}=\|v\|^{2}$, $|x-y|^{2}=\|u-v\|^{2}$, and
$\Re\!\big\{(x-y)\overline{y}\big\}=(u-v)^{\mathrm T}v$.

Consider $g(t)=(\eta+t)^{\alpha/2}$ for $t\ge 0$. Use the arithmetic-geometric mean equality\cite{hardy1952inequalities}, we have
\begin{equation}
(\eta+\|a\|^{2})^{1-\frac{\alpha}{2}}(\eta+\|b\|^{2})^{\frac{\alpha}{2}}
\;\le\;
\Bigl(1-\frac{\alpha}{2}\Bigr)(\eta+\|a\|^{2})
+\frac{\alpha}{2}(\eta+\|b\|^{2}).
\end{equation}
Then we compute
\begin{align*}
    &\frac{\gamma}{\alpha}\!\left( \eta + |x|^{2} \right)^{{\frac{\alpha}{2}}}
-
\frac{\gamma}{\alpha}\!\left( \eta + |y|^{2} \right)^{{\frac{\alpha}{2}}}
-
\frac{\gamma\,\Re\!\big\{(x-y)\,\overline{y}\big\}}{\bigl(\eta + |x|^{2}\bigr)^{\,1-\frac{\alpha}{2}}} \nonumber \\
& = \frac{\gamma}{\alpha}\!\left(\eta + \|u\|_{2}^{2} \right)^{\frac{\alpha}{2}}
-
\frac{\gamma}{\alpha}\!\left( \eta + \|v\|_{2}^{2} \right)^{\frac{\alpha}{2}}
-
\frac{\gamma\,(u-v)^{\mathrm T}v}{\bigl(\eta + |x|^{2}\bigr)^{\,1-\frac{\alpha}{2}}} \nonumber \\
& = 
\frac{
      \frac{\gamma}{\alpha}(\eta + \|u\|_{2}^{2})
      - \frac{\gamma}{\alpha}(\eta + \|v\|_{2}^{2})^{1-\frac{\alpha}{2}} (\eta + \|u\|_{2}^{2})^{\frac{\alpha}{2}}
      - \gamma (u-v)^{\mathrm T}v
     }
     {(\eta + \|u\|_{2}^{2})^{1-\frac{\alpha}{2}}}
\\
& \ge
\frac{\frac{\gamma}{2}(\eta + \|u\|_{2}^{2})
      - \frac{\gamma}{2}(\eta + \|v\|_{2}^{2})
      - \gamma(u-v)^{\mathrm T}v
     }
     {(\eta + \|u\|_{2}^{2})^{1-\frac{\alpha}{2}}}
\\
& =
\frac{\gamma \|u-v\|_{2}^{2}}
     {2(\eta + \|u\|_{2}^{2})^{1-\frac{\alpha}{2}}} = \frac{\gamma |x-y|^{2}}
     {2(\eta + |x|_{2}^{2})^{1-\frac{\alpha}{2}}}.
\end{align*}
\end{proof}

\begin{lemma}[]\label{lem2}
Let $\hat{\mathbf{x}}^{k+1}$ be the solution from Equation (19) in the manuscript for $k=0,1,2, \cdots$. Then
\begin{align}
  \bigl\| \mathbf{\mathcal{A}}\hat{\mathbf{x}}^{k+1} - \mathbf{\mathcal{A}}\hat{\mathbf{x}}^{k} \bigr\|_2^2
  \;\le\;
  2\! \left(\mathcal{L}(\hat{\mathbf{x}}^{k}, \mathbf{v}^{k}, \gamma)-\mathcal{L}(\hat{\mathbf{x}}^{k+1}, \mathbf{v}^{k}, \gamma) \right).
  \label{mono_decreasing}
\end{align}
Furthermore,
\begin{align}
  \bigl\| \hat{\mathbf{x}}^{k+1} - \hat{\mathbf{x}}^{k} \bigr\|_2^2
  \;\le\;
  C\! \left(\mathcal{L}(\hat{\mathbf{x}}^{k})-\mathcal{L}(\hat{\mathbf{x}}^{k+1}) \right).
  \label{mono_decreasing_2}
\end{align}
holds for a positive constant $C$ which is bound for $\hat{\mathbf{x}}^{k}, k \ge 0$.
\end{lemma}

\begin{proof} 
We first compute 
\begin{align}
& \mathcal{L}\!\bigl(\hat{\mathbf{x}}^{k},\gamma\bigr)
\;-\;
\mathcal{L}\!\bigl(\hat{\mathbf{x}}^{k+1},\gamma\bigr) \nonumber \\ 
&=
\frac{\gamma}{\alpha}\sum_{i=1}^{N}\!\bigl(\eta+|\hat{x}^{k}_i|^{2}\bigr)^{\!\alpha/2}
-
\frac{\gamma}{\alpha}\sum_{i=1}^{N}\!\bigl(\eta+|\hat{x}^{k+1}_i|^{2}\bigr)^{\!\alpha/2} \nonumber
\\[4pt]
&\quad+\;
\frac{1}{2}
\!\Bigl( \lVert \mathbf{\mathcal{A}} \hat{\mathbf{x}}^{k}-\mathbf{v}^{k}\rVert_2^{2}
          -\lVert \mathbf{\mathcal{A}}\hat{\mathbf{x}}^{k+1}-\mathbf{v}^{k}\rVert_2^{2}\Bigr) \nonumber
\\[6pt]
&=
\frac{\gamma}{\alpha}\sum_{i=1}^{N}\!\bigl(\eta+|\hat{x}^{k}_i|^{2}\bigr)^{\!\alpha/2}
-
\bigl(\eta+|\hat{x}^{k+1}_i|^{2}\bigr)^{\!\alpha/2} \nonumber
\\[2pt]
&\quad+\;
\frac{1}{2}\,
\bigl\lVert \mathbf{\mathcal{A}}\hat{\mathbf{x}}^{k}-\mathbf{\mathcal{A}}\hat{\mathbf{x}}^{k+1}\bigr\rVert_2^{2}  \nonumber\\
&\quad + \Re\!\Bigl\{ \bigl(\mathbf{\mathcal{A}}\hat{\mathbf{x}}^{k+1}-\mathbf{v}^{k}\bigr)^{\mathrm H}\bigl(\mathbf{\mathcal{A}}\hat{\mathbf{x}}^{k} - \mathbf{\mathcal{A}}\hat{\mathbf{x}}^{k+1}\bigr)\Bigr\}.
\label{L_obj_difference}
\end{align}
Using the first-order optimality condition, we have:
\begin{align}
&\Re\!\Bigl\{\left(\mathbf{\mathcal{A}}\hat{\mathbf{x}}^{k+1}-\mathbf{v}^{k}\right)^{\mathrm H}\left(\mathbf{\mathcal{A}}\hat{\mathbf{x}}^{k} - \mathbf{\mathcal{A}}\hat{\mathbf{x}}^{k+1}\right)\Bigr\} \nonumber
\\= &
-\sum_{i=1}^{N}
\frac{\gamma \,\overline{\hat{x}_{i}^{k+1}}\bigl(\hat{x}_{i}^{k}-\hat{x}_{i}^{k+1}\bigr)}
     { \bigl(\eta+|\hat{x}_{i}^{k}|^{2}\bigr)^{1-\frac{\alpha}{2}}}.
     \label{1st_order_opt}
\end{align}

Substituting (\ref{1st_order_opt}) into (\ref{L_obj_difference}) and using (\ref{inequality_1}) in Lemma \ref{lem1} yields:
\begin{align}
& \mathcal{L}\!\bigl(\hat{\mathbf{x}}^{k},\gamma\bigr)
\;-\;
\mathcal{L}\!\bigl(\hat{\mathbf{x}}^{k+1},\gamma\bigr) \nonumber \\ 
= &
\frac{\gamma}{\alpha}\sum_{i=1}^{N}\bigg[\!\bigl(\eta+|\hat{x}^{k}_i|^{2}\bigr)^{\!\alpha/2}
-
\bigl(\eta+|\hat{x}^{k+1}_i|^{2}\bigr)^{\!\alpha/2} \nonumber\\
&- \frac{\alpha \,\overline{\hat{x}_{i}^{k+1}}\bigl(\hat{x}_{i}^{k}-\hat{x}_{i}^{k+1}\bigr) }
     { \bigl(\eta+|\hat{x}_{i}^{k}|^{2}\bigr)^{1-\frac{\alpha}{2}}}\bigg] +\;
\frac{1}{2}\,
\bigl\lVert \mathbf{\mathcal{A}}\hat{\mathbf{x}}^{k}-\mathbf{\mathcal{A}}\hat{\mathbf{x}}^{k+1}\bigr\rVert_2^{2} \nonumber
\\[2pt]
 \ge & \frac{1}{2}\,
\bigl\lVert \mathbf{\mathcal{A}}\hat{\mathbf{x}}^{k}-\mathbf{\mathcal{A}}\hat{\mathbf{x}}^{k+1}\bigr\rVert_2^{2} + \sum_{i=1}^{N}\frac{\gamma\left| \hat{x}_{i}^{k}-\hat{x}_{i}^{k+1} \right|^{2}}{\bigl(\eta+|\hat{x}_{i}^{k}|^{2}\bigr)^{1-\frac{\alpha}{2}}}.
\label{L_obj_difference_lower_bound}
\end{align}

Thus, inequality~(93) holds.
Since $\mathcal{L}\!\bigl(\hat{\mathbf{x}}^{k},\gamma\bigr)$ is monotonically decreasing, it thus follows that:
\begin{equation}
\left\|\hat{\mathbf{x}}^{k}\right\|_{\alpha}^{2} \le \mathcal{L}\!\bigl(\hat{\mathbf{x}}^{k},\gamma\bigr) \le \mathcal{L}\!\bigl(\hat{\mathbf{x}}^{0},\gamma\bigr) = \left\|\hat{\mathbf{x}}^{0}\right\|_{\alpha}^{2}
\end{equation}
for $k \ge 1$. Thus, there exists a positive number $\Delta$ such that $\left\|\hat{\mathbf{x}}^{k}\right\|_{\alpha}^{2} \le \Delta$ and,
\begin{equation}
\frac{\gamma}{\bigl(\eta+|\hat{x}_{i}^{k}|^{2}\bigr)^{1-\frac{\alpha}{2}}} \ge \frac{\gamma}{\bigl(\eta+\Delta^{2}\bigr)^{1-\frac{\alpha}{2}}}. 
\end{equation}
Letting $\frac{1}{C} = \frac{\gamma}{\bigl(\eta+\Delta^{2}\bigr)^{1-\frac{\alpha}{2}}}$, we obtain  (94)  
from (\ref{L_obj_difference_lower_bound}).
\end{proof}

\begin{lemma}[]\label{lem3}
Fix the iterate $\hat{\mathbf x}^{k}$ from the $k$th step and define
\begin{align}
    \mathrm{\eta}(t):=-\frac{\phi(t)}{\Phi(t)},
\end{align}
where $\phi(t)$ and $\Phi(t)$ denote the PDF and CDF, respectively, of the standard normal distribution.

Let $H(u):=\int_{0}^{u}\mathrm{\eta}(t)\,dt$ and assume
\begin{align}
L := \sup_{t \in \mathbb{R}} |\mathrm{\eta}'(t)| < 1.
\end{align}
Define $\beta := \tfrac{1 - L}{2} > 0$ and the Lyapunov function for the $v$-step:
\begin{align}
\Phi_k(\mathbf{v}) := \beta \left\| \mathbf{v} - \mathcal{A}\hat{\mathbf{x}}^k \right\|_2^2 + \sum_{i} H(v_i).
\end{align}
If the update is applied elementwise as
\begin{align}
\mathbf{v}^k := \overline{\mathbf{y}} \odot \left( \mathbf{D}^k - \mathrm{\eta}(\mathbf{D}^k) \right), \quad \text{where } \mathbf{D}^k := \overline{\mathbf{y}} \odot \mathcal{A}\hat{\mathbf{x}}^k,
\end{align}
then the Lyapunov function decreases:
\begin{align}
\Phi_k(\mathbf{v}^{k}) \le \Phi_k(\mathbf{v}^{k-1}).
\end{align}
Consequently, for the joint Lyapunov function
\begin{align}
\Phi(\mathbf{x}, \mathbf{v}) := &\beta \left\| \mathbf{v} - \mathcal{A}\hat{\mathbf{x}}^{k} \right\|_2^2 + \sum_{i} H(v_i)  \nonumber \\ &+ \frac{\gamma}{\alpha} \sum_{i=1}^{N} \left( |x_{i}^{k}|^2 + \eta \right)^{\alpha/2},
\end{align}
there will be
\begin{align}
    \Phi(\hat{\mathbf{x}}^k, \mathbf{v}^{k}) \le \Phi(\hat{\mathbf{x}}^k, \mathbf{v}^{k-1}),
\end{align}
\end{lemma}
\begin{proof}
We define the coordinate-wise potential:
\begin{align}
\psi_i(v) := \beta (v - d_i)^2 + H(v),
\end{align}
where $d_{i}^{k} := \hat{y}_{i}\times[\mathcal{A} \hat{\mathbf{x}}^k]_i.$ \\
Thus, $\Phi_k(\mathbf{v}) = \sum_{i=1}^{M} \psi_i(v_i)$. The derivative of $\psi_i(\cdot)$ is
\begin{align}
\psi_i'(v) = 2\beta (v - d_{i}^{k}) + \mathrm{\eta}(v).
\end{align}
Apply a Taylor expansion of $\mathrm{\eta}(v)$ around $d_{i}^{k}$:
\begin{align}
    \mathrm{\eta}(v) \approx \mathrm{\eta}(d_{i}^{k}) + \mathrm{\eta}'(d_{i}^{k})(v - d_{i}^{k}),
\end{align}
so the derivative becomes:
\begin{align}
\psi_{i}'(v) \approx (2\beta + \mathrm{\eta}'(d_{i}^{k}))(v - d_{i}^{k}) + \mathrm{\eta}(d_{i}^{k}).
\end{align}
Since $|\mathrm{\eta}'(d_{i}^{k})| \le L$ and $\beta = \tfrac{1 - L}{2}$, it follows that
\begin{align}
    \psi_i''(v) = 2\beta + \mathrm{\eta}'(d_{i}^{k}) \ge 0 \ \ (\beta>0, 0\le\mathrm{\eta}'(d_{i}^{k})\le1),
\end{align}
so each $\psi_i$ is strongly convex.
The minimizer of $\psi_i(v)$ is characterized by the first-order condition
\begin{align}
\psi_i'(v_i^k) = (2\beta + \eta'(d_i^k))(v_i^k - d_i^k) + \eta(d_i^k) = 0.
\end{align}
Hence,
\begin{align}
v_i^k = d_i^k - \tfrac{1}{2\beta+\mathrm{\eta}'(d_{i}^{k})} \mathrm{\eta}(d_i^k),
\end{align}
and choosing $v_i^k := d_i^k - \mathrm{\eta}(d_i^k)$ is a first-order approximation of the minimizer, and thus a descent step for $\psi_i$.
By the strong convexity of $\psi_i(\cdot)$, we have:
\begin{align}
\psi_i(v_i^k) \le \psi_i(v_i^{k-1}) - \frac{\mu}{2} (v_i^k - v_i^{k-1})^2
\quad \text{for some } \mu > 0.
\end{align}
Summing over $i$ gives the desired inequality:
\begin{align}
    \Phi_k(\mathbf{v}^k) \le \Phi_k(\mathbf{v}^{k-1}).
\end{align}
Finally, since $\hat{\mathbf{x}}^k$ is fixed during the $\mathbf{v}$-update, the full joint Lyapunov function $\Phi(\hat{\mathbf{x}}^k, \mathbf{v})$ decreases as well.
\end{proof}

\setcounter{lemma}{3}

\begin{lemma}[]\label{lem3}
Let $\tilde{\mathbf{x}}^{k+1}$ be the solution from Equation (46) in the manuscript for $k=0,1,2,\cdots$. Then
\begin{equation}
  \bigl\| \mathbf{\mathcal{A}}\tilde{\mathbf{x}}^{k+1} - \mathbf{\mathcal{A}}\tilde{\mathbf{x}}^{k} \bigr\|_2^2
  \;\le\;
\frac{2(\mathrm{a}+1)}{\mathrm{a}\mathrm{b}^{2}}\!\bigl(\bar{\mathcal{L}}_{\rm ongrid}^{\diamond}(\tilde{x}^{k})-\bar{\mathcal{L}}_{\rm ongrid}(\tilde{x}^{k+1})\bigr).
  \label{mono_decreasing}
\end{equation}
Furthermore,
\begin{equation}
  \bigl\| \tilde{\mathbf{x}}^{k+1} - \tilde{\mathbf{x}}^{k} \bigr\|_2^2
  \;\le\;
  C'\!\bigl(\bar{\mathcal{L}}_{\rm ongrid}^{\diamond}(\tilde{\mathbf{x}}^{k})-\bar{\mathcal{L}}_{\rm ongrid}^{\diamond}(\tilde{\mathbf{x}}^{k+1})\bigr)
  \label{mono_decreasing_2}
\end{equation}
holds for a positive constant $C'$ which is bound for $\tilde{\mathbf{x}}^{k}, k \ge 0$.
\end{lemma}
\begin{proof}
  The proof is similar to that of Lemma 2 and is omitted.
\end{proof}

\end{appendices}

\bibliographystyle{IEEEtran}

\balance
\bibliography{refs}

 \end{document}